# Experimental Simulations of Shock Textures in BCC Iron: Implications for Iron Meteorites

Eiji Ohtani[1]

Corresponding author

Email: eohtani@tohoku.ac.jp

Toru Sakurabayashi[1],

Email: delpieroo552@gmail.com

Kosuke Kurosawa[2]

E-mail: kosuke.kurosawa@perc.it-chiba.ac.jp

[1] Department of Earth Science, Graduate School of Science, Tohoku University, Sendai 980-8578, Japan

[2] Planetary Exploration Research Center, Chiba Institute of Technology, Chiba 275-0016, Japan

## Abstract

Neumann band in iron meteorites, which is deformation twins in kamacite (Fe–Ni alloy), has been known to be a characteristic texture indicating ancient collisions on parent bodies of meteorites. We conducted a series of shock recovery experiments on bcc iron with the projectile velocity at 1.5 km/sec at various initial temperatures, room temperature, 670 K, and 1100 K, and conducted an annealing experiment on the shocked iron. We also conducted numerical simulations with the iSALE-2D code to investigate peak pressure and temperature distributions in the nontransparent targets. The effects of pressure and temperature on the formation and disappearance of the twins (Neumann band) were explored based on laboratory and numerical experiments. The twin was formed in the run products of the experiments conducted at room temperature and 670 K, whereas it was not observed in the run product formed by the impact at 1100 K. The present experiments combined with the numerical simulations revealed that the twin was formed by impacts with various shock pressures from 1.5–2 GPa to around 13 GPa. The twin in iron almost disappeared by annealing at 1070 K. The iron meteorites with Neumann bands were shocked at this pressure range and temperatures at least up to 670 K, and were not heated to the temperatures above 1070 K after the Neumann band formation.

# 1 Introduction

Wasson (1967) and Wasson et al. (1998) analyzed 700 iron meteorites, initially for Ni, Ga, Ge, and Ir and later for Cr, Co, Cu, As, Sb, W, Re, Pt, and Au, using instrumental neutron activation analyses. Based on these analyses, iron meteorites have been divided into 13 groups according to their chemical compositions (Goldstein et al., 2009). The Neumann band has been observed in kamacite (body-centered-cubic [bcc] phase) separated as the Widmanstätten pattern in most groups of iron meteorites such as Goose Lake IAB (Scott, 2020) and IVA, IVB, and ungrouped iron meteorites such as Hammond and Babb's Mill (Blake's Iron and Troost's Iron)

(Yang et al. 2011).

Both carbonaceous chondritic (CC)-type and non-carbonaceous (NC)-type iron meteorites (Hilton et al., 2019; Scott, 2020) fractionated in their parent bodies in the early stage of formation of the solar system, perhaps within 1–3 My after the formation of calcium-aluminum-rich inclusions in the solar nebula (Kruijer et al., 2017; Hilton et al., 2019). Such early metal–silicate fractionation occurred perhaps because of melting caused by the radiogenic heating by extinct radiogenic elements such as $^{26}$Al (e.g., Goldstein et al., 2009). Therefore, iron meteorites possess information on their early differentiated planetesimals.

Many iron meteorites, however, have experienced textural deformation because of mutual collisions, which would modify the original information. Therefore, if we accurately understand the impact conditions required for producing the shock features in iron meteorites, we would be able to decode the impact environment in the early solar system.

Although magnitudes of the shock events in ordinary chondrites have been classified as the shock stage from S1 to S6 based on their shock textures (Stöffler et al. 1991), there are few systematic classifications of shock stages in iron meteorites. Yang et al. (2011) made an effort to classify the four shock stages for IVA, IVB, and some ungrouped iron meteorites based on various shock textures. According to them, the Neumann band is a prominent signature of stage 1, which shows the records of relatively weak collisions, and it can be removed by annealing caused by the later heating events. Therefore, the formation and disappearing conditions of the Neumann band are useful to identify unaltered or less-altered iron meteorites after their formation in the early solar system. Because of such importance, we made experimental simulations of the formation and disappearance of the Neumann band.

The Neumann band has been known to be a deformation twin (Uhlig, 1955) formed by the intracrystalline plastic deformation in many metals. This deformation twin in bcc iron shows the twin boundary of (112) and moving directions of atoms along [111] on the twin boundaries (Calister and Rethwisch, 2000；Murr et al., 2002a). It has been considered that the Neumann band in iron meteorites was formed by mutual collisions between planetesimals (Bischoff and Stöffler, 1992).

Murr et al. (2002a, b) conducted impact experiments with bcc iron and SUS304 (an austenitic stainless steel), and observed deformation twins in the recovered iron samples. Their shock experiments reported velocities at 0.5–3.8 km/sec, and measured the depths where twins formed. Based on the relations of the impact velocity and the depths of the twin formation, they estimated that an impact velocity greater than 0.1 km/sec (equivalent to 1 GPa) is needed for producing twins (Murr et al., 2002b). Although impact experiments on iron–nickel alloys at low temperatures down to 110 K (Marchi et al., 2020) and those of iron at 76–573 K (Rohde, 1969) have been conducted previously, these experiments did not make detailed analyses pertaining to twin formation and disappearance.

In this study, we conducted laboratory and numerical experiments with bcc iron targets at a variety of initial temperatures to investigate the structural changes of iron against a propagation of compressive pulse. We reproduced deformation twins experimentally in bcc iron and revealed the temperature and pressure dependence of the twin formation. We also conducted an annealing experiment on iron having twins to explore the temperature required for their disappearance. We discussed the formation and disappearance conditions of Neumann bands in shocked iron meteorites based on our experiments and numerical simulations.

## 2 Experimental Methods

We conducted shock recovery experiments with a light gas gun. An iron projectile was accelerated onto an iron target. We observed the shock textures including deformation twins in the recovered samples. Here, we summarize the shock apparatus, heating system to realize the high-temperature impact experiments, and procedure of analyses of the samples recovered from the impact experiments. We also conducted numerical simulations to estimate the pressure and temperature distributions in iron targets under the same impact conditions archived in the laboratory experiment.

**2.1 Two-stage light gas gun used for the impact experiments**

The two-stage light gas gun installed in the Institute of Fluid Science of Tohoku University, Japan, was used to accelerate a projectile. The instrument is schematically shown in Figure 1. Helium gas was compressed by a piston made of high-density polyethylene (33 mm in diameter and 60 mm in length). A mylar film (500 μm in thickness) was used as the second diaphragm to confine the gas pressure. The pistons before and after the experiment are shown in Supplementary Figure S1. We generated the projectile velocity of 1.5 km/sec in the present experiments. The detailed information on the two-stage light gas gun is given elsewhere (e.g., Shinohara, 2002). We employed the laser light blocking method to measure the projectile velocity, $V$, and the impedance matching method to calculate the pressure at the epicenter (Miller et al., 1991; Ahrens, 1987). The projectile velocity was determined from the time of the flight between two laser beams with the interval of 520 mm. We employed a GaAs semiconductor laser with a power of 3.2 mW (Kikoh Giken Co., Ltd., Hyogo, Japan). The error in velocity estimation was 1% or less. We present each experimental condition including the impact velocity in Table 1.

**2.2 Projectile and target material**

The projectile and target samples were iron with the purity of 99.9 %. The projectile used in the present experiment is shown in Supplementary Figure S2 and Table 1. The projectile body was made of polycarbonate with 10 mm in diameter and 13.5 mm in length. A bcc iron disc with 8 mm in diameter and 3.5 mm in height was placed at the top of the polycarbonate body. The total weight of the projectile was 2.5 g.

Bcc iron was also used as the target. The samples were shaped into cylinders with a diameter of 38 mm and a height of 20 mm or 30 mm. The longer target was used in the experiments at high temperatures (670 K and 1100 K). The iron sample was placed in a molybdenum capsule (50 mm in outer diameter, 38 mm in inner diameter, depth of 8 mm, and height of 10 mm) to heat the iron sample effectively using the induction furnace. The target iron sample embedded in a molybdenum capsule is shown in Supplementary Figure S3.

## 2.3 Heating and the temperature measurement of the sample

We heated the target samples before the impact of the projectile to investigate the effect of initial temperature on the textures of the shocked samples. We used an induction furnace (TRD-02001, Dai-ichi Kiden Co. Ltd., Tokyo, Japan) for heating the target samples. A schematic image of the sample chamber with the induction furnace and the temperature measurement system is shown in Figure 2. The induction coil of the furnace was covered by a ceramic brick cylinder with an outer diameter of 70 mm and an inner diameter of 50 mm (as shown in Figure 2) to avoid damaging the coil from the fragments of the projectile and target during impact. The target iron sample and induction furnace were placed in the sample chamber of the two-stage light gas gun. The sample chamber was evacuated using a rotary vacuum pump to 20 Pa, which is within the suitable condition of heating by the induction furnace (667 Pa). The temperature of the sample surface was measured by radiation from the sample using an optical pyrometer (IR-CAS2TN, CHINO Co. Ltd., Tokyo, Japan). We measured the irradiance at 0.9 μm using the silicon detector for temperature measurement. The uncertainty of the sample temperature measurement was ±5 °C below 1000 °C. The image of the heated sample and the temperature reading of the optical pyrometer on the sample surface at the time of collision were recorded by a video camera (DM-FVM30, Canon, Co. Ltd, Tokyo, Japan). The heating rate was about 7 K/min, and the cooling rate by shutting off the electrical supply was 20 K/min. The two-stage gas gun was operated within one minute after shutting off the electrical supply for heating for safety. Thus, the uncertainty of the experimental temperature was within 20 K. The sample temperature is assumed to be homogeneous and the same as that on the surface because of heating by the induction furnace and high thermal conductivity of bcc iron, which is shown in the literature to range from 23 mm$^2$/sec at 300 K to 5 mm$^2$/sec at 1000 K (Monaghan and Quested, 2001).

## 2.4 Analysis of the recovered samples

Figure 3 shows the backscattered electron (BSE) image (a) and secondary electron image (SEI) (b) of the texture of the etched bcc iron target material. The grain size of the starting material was 100–200 μm in diameter. Craters were formed in the iron targets after the impact experiments as shown in Figure 4. To observe the shocked texture of the recovered sample, we cut the sample into two pieces and polished it with silicon carbide and fine diamond powders making a mirror surface of the cross section across the center of the crater. We etched the polished surface with Nital for a few seconds to perform textural observation of the polished cross section. The textual observations were conducted using a reflective optical microscope (Nikon PTIPHOT2-POL) and a scanning electron microscope (JEOL JSM5410).

We determined the number of twins in the rectangle region with the size of 400 μm × 300

μm at different depths of the polished samples measured from the crater surface of the target. We also measured the mean grain sizes of the samples and obtained the number density of twins in individual grain (number per unit area, $1/cm^2$) as shown in Table 2. We did not measure the grain orientations in the recovered samples in this measurement.

### 2.5 Shock physics modeling using iSALE-2D code

We conducted two-dimensional impact simulations using the iSALE shock physics code (Amsden et al., 1980; Ivanov et al., 1997; Wünnemann et al., 2006) to estimate the pressure and temperature distributions in the nontransparent samples during the laboratory experiments. We used the two-dimensional version of the iSALE, which is referred to as iSALE-Dellen (Collins et al., 2016). A cylindrical coordinate was employed to model vertical impacts performed in the experiments. The analytical equations of state (ANEOS) (Thompson and Lauson, 1972) pertaining to iron (Thompson, 1990) were used for both the iron projectile and target. To treat the elastoplastic behavior of the shocked iron, we employed the constitutive equations in the simulations. We used the Johnson–Cook strength model (JNCK) with the parameters pertaining to ARMCO-iron (Johnson and Cook, 1983) because the relation between stress and strain for ARMCO-iron has been well understood (e.g., Johnson and Cook, 1983) and the ARMCO-iron contains a small fraction of impurities, which is less than 1%. The JNCK parameters and calculation settings used in the calculations are listed in Tables 3 and 4, respectively.

## 3 Results

### 3.1 The deformation twins created by the impact

We conducted three shock experiments at room temperature (Run 011), 1100 K (Run 012),

and 670 K (Run 013). The projectile collided with the sample at a velocity of 1.53($\pm$0.02) km/sec at room temperature (Run 011). Figure 4(a) shows the crater and its polished cross section. The crater with a diameter of 17.8 mm and a depth of 6.8mm was formed by the impact under this condition. Figure 5 shows the BSE images of the texture of the polished section and the SEI of the highly magnified textures of different depths near the crater. A composite BSE image of the whole sample is shown in Figure 6. In the region from the surface down to 1.0-mm depth, we observed concentric flow textures of grains as shown in Figure 5. We observed linear textures with 20 μm length and 0.5 μm width in this region. In the region from 1.0- to 2.6-mm depths, we observed twins together with flow textures. In the deeper region from 2.6 mm to 15.5 mm, we only observed twins (see Figures 5 and 6). The number of twins decreases with depth in this sample.

In the experiment conducted at 1100 K (Run 012), we used a 30-mm-long sample to investigate the temperature effect on the twin formation. The temperature of the sample was increased to 1110 K first, then we turned off the electric power of the induction furnace, and the projectile was accelerated to a velocity of 1.51($\pm$0.02) km/sec. Because the cooling rate of the sample was slow (20 K/min as mentioned), the sample temperature at the time of the collision was estimated to be 1090–1110 K in this experiment. Figure 4(b) shows the polished section of the recovered sample. A 22.4-mm-wide and 10.1-mm-deep crater was formed by the impact at 1100 ± 10 K. Figure 7 shows the BSE image of the texture of the whole sample. The polished textures of this sample (Run 012) were remarkably different from those of the shocked sample at room temperature (Run 011). We did not find any flow textures but observed polygonal crystals with a grain size of about 70 μm from beneath the crater down to a depth of 2.6 mm. We also observed polygonal crystals with a larger grain size of 100–200 μm in the deeper regions, and we did not identify any twins to the rear surface of the sample.

In the third experiment (Run 013), the projectile was collided with a 30-mm-long sample at the velocity of 1.47($\pm$0.02) km/sec at 670 K. The recovered sample and its cross section are shown in Figure 4(c). The diameter and the depth of the crater were 16.2 mm and 6.5 mm, respectively. The BSE images of the whole views of the polished section, and magnified local SEI images are shown in Figure 8. Grains with several tens μm in diameter were observed in the sample at the depths of 1.2 mm. Both the flow textures and twins were observed at depths from 1.2 mm to 2.5 mm. We observed only twins at depths from 2.5 mm to around 20 mm of the sample, as shown in Figure 8. The number of twins decreased with the depths of the sample. The craters have slightly asymmetrical shapes, as shown in Figure 4. Although the cause of asymmetry is not well known, it could be because of a slightly tilted collision of the projectile.

We conducted an annealing experiment on the shocked target at room temperature, which has the deformation twins. The recovered sample from Run 011 was shaped into a plate with the size of 18 mm height×2 mm width×1 mm thickness. The plate was heated for ten minutes at 1070 K, which is below the bcc-fcc phase boundary (1123 K) and was cooled to room temperature with a cooling rate of 20 K/min. The BSE and SEI images of the whole view of the plate after annealing at 1070 K (Run 011H) are shown in Figure 9. Although we observed the oxidized layer with a thickness around 50 μm on the surface of the annealed sample, we did not observe the oxidized layer in the deeper interior of the heated sample. We observed remarkable changes in textures after annealing at 1070 K for ten minutes. The flow patterns formed near the crater by the room temperature impact (Run 011) disappeared to form polygonal grains with a size of 40–100 μm (Run 011H).

### 3.2. Impact simulation using iSALE-2D code

We conducted the impact simulations under the same conditions, including the sample temperatures, using iSALE-2D code. The distributions of the peak pressure and temperature during the impacts at room temperature, 1070 K, and 670 K are shown in Figure 10. The depth profiles of the peak pressure and temperature along the central axis of the iron targets of these impacts are shown in Figure 11. An increase in pressure to 30 GPa was observed within the 5-mm depths from the craters for all the runs, whereas the pressure decreased gradually from 10 GPa to less than 1 GPa at the depths greater than 5 mm to the bottom of the target samples. The temperature profiles also showed similar distributions, as shown in the bottom column of the figure. A rapid increase in temperature appeared at depths shallower than 5 mm, whereas the temperature increase was limited below 160 K in the region deeper than 5 mm depth, and the temperature increase was negligible from the 10 mm depth to the bottom of the target samples. The present simulation is consistent with the shock pressure of 30 GPa estimated by the shock Hugoniot (Bischoff and Stöffler, 1992; Mur et al., 2002a) at the top of the crater within the 5-mm depths; however, the pressure decayed continuously to below 2 GPa at the bottom of the sample. Therefore, it is important to consider this decay of the maximum pressure when discussing the formation conditions of twins.

### 3.3 Distribution of deformation twins in iron target samples

Variations in the density of twins within the target samples for three experiments (Runs 011, 013, and 011H) are summarized in Figure 12. Because we did not observe any twins in the shocked target at 1100 K (Run 012), the data obtained only at room temperature and 670 K are shown. The absence of the twins in this sample is consistent with the result of the annealing experiment.

The distribution of twins was expressed as their density in a unit area (Number/cm$^2$) in one grain of the sample following the definition of the dislocation density in crystals and is shown in Table 2. The twin was not observed near the crater surface to 5-mm depths perhaps because of an intensive deformation associated with the pressure and temperature increase to 30 GPa and above 800 K, respectively, in the impacts at room temperature and 670 K. No significant difference in the number density profiles with pressure was observed between the samples collided at room temperature (Run 011) and 670 K, as shown in Figure 12(b). The distribution of twns annealed at 1070 K below the bcc-fcc transition for 10 min (Run 011H) shows a remarkable reduction of the number density at all depths, as shown in Figure 12(a).

## 4 Discussion

4.1. Conditions for twin formation combining impact experiments and simulations.

The twins were formed even near the rear surfaces of the target samples where the shock pressures decayed below 2 GPa estimated by the iSALE (Figure 11a, c), suggesting that the threshold pressure of the twin formation is close to 1.5–2 GPa. Murr et al. (2002b) reported the relation of the impact velocity and the depth of twin formation and estimated the threshold impact velocity for the twin formation to be 0.1 km/sec (corresponding to 1 GPa) at room temperature, which is consistent with our observations.

We can estimate the thermal history of iron meteorites experienced during and after collision based on their shock textures. Our experiments revealed that the Neumann band is a signature of shock above 1.5–2 GPa, which can be formed by impacts below temperatures about 670 K. However, our result also showed that too high degrees of compression above 13 GPa cannot produce the twins possibly because of too high dislocation density in one grain.

Therefore, the Neumann band can be used as a shock indicator between 1.5–2 GPa and around 13 GPa. The relations between the twin density and pressure are also presented as shown in Figure 12(b). We found that the twin density can be well fitted by a simple linear relation as a function of peak pressure. The phase boundary between bcc and hcp iron locates at around 13 GPa. Figures-11 and 12 also suggest saturation of the twin density due to large deformation. Although we observed the texture after pressure release, the recent in-situ XRD measurements of the released stages from shocked iron using XFEL (Hwang et al. 2020) suggested the process occurred in the large deformation region of our recovered samples, i.e., the phase change occurred very quickly and even fcc phase could be formed due to expansion of the samples. These processes likely occurred in the highly deformed regions of our samples.

We found that the initial temperature does not largely affect the twin density at least up to 670 K, indicating that the production rate of twins is high enough to neglect the effects of annealing kinetics at least below 670 K. The present experiments combined with the numerical simulations revealed that the Neumann band were shocked by impacts with shock pressures from 1.5–2 GPa to around 13 GPa and temperatures at least up to 670 K.

4.2. Annealing and disappearance of twins

Our annealing experiment (Run 011H) revealed that twins disappeared easily in bcc iron by annealing and recrystallization at 1070 K below the bcc-fcc transition temperature only for ten minutes. This observation is consistent with experimental results of high-temperature annealing of iron meteorites conducted by Jain and Lipschutz (1968), Buchwald (1975), and Davidson (1940). These previous annealing experiments on iron meteorites indicated that the samples recrystallized at temperatures above 870 K, which is below the bcc-fcc phase transition boundary, after 0.5–3 hours of annealing. Annealing experiments on heavily deformed iron,

Fe–0.3 wt%Al, and Fe–0.3 wt%Si alloys also indicated that recrystallization started at temperatures above 770 K with a heating rate of 10 K/min and significant grain growth occurred at 1070 K (Tomita et al., 2017), which is also consistent with our observation of twin disappearance in the shocked iron sample.

Recrystallization including reduction of dislocation density has been studied extensively (Humphreys et al., 2017). The conventional analysis for recrystallization may be applied to reduce the twin density by recrystallization, and the decrease rate may be expressed by an empirical $n$-th order recrystallization kinetics expression, $d\rho/dt = -k\rho^n$, where $k$ is the rate constant for recovery (Humphreys et al., 2017), and $\rho$ is the twin density. Different values are proposed for $n$-values depending on the materials; $n = 2$ for the recovery of olivine (Farla et al., 2010) and LiF (Li, 1966), $n = 2$ (Van Drunen and Saimoto, 1971) or 3 (Prinz et al., 1982) for Cu, and $n = 3$ for Ni (Prinz et a., 1982). Therefore, we adopted $n = 2$ and 3 for decrease in twin density by annealing. By fitting our data by this equation, $d\rho/dt$ can be obtained with a parameter $k = 3.084 \times 10^{-8}$ for $n = 2$ and $k = 4.316 \times 10^{-13}$ for $n = 3$ at 1070 K for 10 min. Based on this equation, we can estimate the time dependency of the decrease in twin density for various durations from 1 min to 1 hour at 1070 K as shown in Figure 13(a). Significant reduction of the twin density occurred by annealing at 1070 K as shown in this figure.

Decrease in twin density because of annealing and recrystallization at different temperatures may be estimated by the temperature dependence of the rate constant, $k$, which can be expressed by $k = k_0 \exp(-E_a/RT)$, where $E_a$ is the activation energy for the process considered, $k_0$ is a constant, $R$ is the gas constant, and $T$ is the temperature. Because we obtained $k(1070\text{ K})$, as mentioned, we can obtain a temperature-dependent rate constant $k(T)$ if $E_a$ is given. The activation energy for the recovery is expected to be close to the self-diffusion of the metals for Cu (Cottrell and Aytekin 1950), Fe (Michalak and Paxton, 1961), and Zn (Van

Drunen and Saimoto, 1971). Experiments on recovery/recrystallization of steel suggest the activation energy, $E_a$, is around 250 kJ/mol, which is close to that of the iron self-diffusion (e.g., Watanabe and Karashima, 1970; Glover and Stellars, 1973). Therefore, the activation energy of 250 kJ/mol was adopted in the recrystallization of twins. Finally, we can estimate the temperature dependence of the disappearance of twins, as shown in Figure 13(b). This figure indicates annealing and disappearance of twin occur above 700 K, which is consistent with annealing experiments on iron meteorites (Jain and Lipschutz, 1968; Buchwald, 1975). This is the reason why the shocked sample at 670 K was free from the annealing. In our impact experiment conducted at 1100 K (Run 012), we cannot observe twins in the iron target, as shown in Figure 7. The cooling rate after the impact at 1100 K is estimated to be 20 K/min as shown in Section 2.2. Therefore, the disappearance of twins at 1100 K impact (Run 012) annealed for 10 min during cooling to 900 K is consistent with the recrystallization rate of twin estimated here.

Our experiments showed that twins were formed by impacts at least up to 670 K, and it was not formed by high-temperature impacts above 1100 K. The annealing experiment and the analysis of the kinetics given in Figures 13 (a) and (b) revealed that the twin in iron disappeared easily by reheating and annealing at 1070 K.

The bcc-fcc transition temperature in iron is 1184 K, whereas the transition of kamacite with 5–7.5 mol% Ni occurs at 770–905 K (Reisener and Goldstein, 2003). Therefore, the maximum temperature for survival of the Neumann band in kamacites might be as low as the temperature of the bcc-fcc equilibrium phase boundary. Although we used iron in this study, our result may provide insight into Neumann band formation in iron meteorites because kamacite can survive even at 1070 K metastably because of sluggish transition kinetics (Dunlop and Özdemir, 2007) when it was reheated rapidly. Therefore, iron meteorites with Neumann bands were not heated to the temperatures above 1070 K after the Neumann band formation.

## 5 Conclusions

A series of shock recovery experiments at three different sample temperatures were conducted using a two-stage light gas gun with an induction furnace. Textures of the recovered iron were studied with the impact velocity of about 1.5 km/sec in the temperature range from room temperature to 1100 K. The twin was observed in the run products of room temperature and 670 K, whereas it was not observed in the run product recovered from the impact at 1100 K.

We conducted numerical simulations of these impacts to estimate the maximum pressure and temperature distributions in the targets during the impacts. We revealed that the twin can be formed at relatively low pressures from 1.5–2 GPa to around 13 GPa at low temperatures at least up to 670 K. The present experiments combined with the numerical simulations revealed that the Neumann band were shocked at this pressure range by impacts and temperatures at least up to 670 K. Our annealing experiments also implied that the twin in iron recrystallizes and disappears easily by heating at around 1070 K. Iron meteorites with Neumann bands were not heated above 1070 K after the Neumann band formation.

## Abbreviations

SUS304: Steel Use Stainless 304, Nital: Ethanol containing 5 vol % of nitric acid, BSE: Backscattered electron, SEI: Secondary electron image, bcc: Body-centered cubic, iSALE: Shock physics code for impact-simplified arbitrary Lagrangian Eulerian, JNCK: Johnson-Cook strength model, ANEOS: Analytical equation of state developed for shock physics, ARMCO-iron: Pure iron produced by American Rolling Mill Company.

# Declarations

The authors have no conflicts of interest.

## Availability of data and material

All data is available in the main text and the supplementary materials. Additional data may be available from the corresponding authors upon reasonable request.

## Competing interests

The authors declare that they have no competing interest.

## Funding

This work was supported by JSPS KAKENHI, JP11874071, JP09304051, and JP20H00187 to EO. KK was supported by JSPS KAKENHI JP18H04464, JP19H00726, and JP21K18660.

## Authors' contributions

E. Ohtani devised the project, designed the experiments, prepared the heating system, and attended the impact experiment. T. Sakurabayashi prepared the samples, conducted the impact experiments, and collected data and analyzed the recovered samples. K. Kurosawa conducted iSALE-2D simulation of the impacts. E. Ohtani and K. Kurosawa wrote the manuscript with inputs from T. Sakurabayashi.

## Authors' information

Eiji Ohtani is an emeritus professor at the Graduate School of Science, Tohoku University. Toru

Sakurabayashi was a graduate student of Tohoku University, and now is working at Nippon Electric Glass Co., Ltd. Kosuke Kurosawa is a senior staff scientist of Planetary Exploration Research Center, Chiba Institute of Technology.

**Acknowledgments**

The authors appreciate W. McDonough for stimulating discussion on this work including the origin of iron meteorites. The authors thank Akio Suzuki of Graduate School of Science, Tohoku University, and Professor (Emeritus) Kazuyoshi Takayama and technical staff of Institute of Fluid Science, Tohoku University for helping them conduct shock experiments and analyze recovered samples. The authors appreciate the developers of iSALE, including G. Collins, K. Wünnemann, B. Ivanov, J. Melosh, and D. Elbeshausen. The authors also thank Tom Davison for the development of the pySALEPlot. This work was supported by JSPS KAKENHI, JP11874071, JP09304051, and JP20H00187 to EO.   KK was supported by JSPS KAKENHI JP18H04464, JP19H00726, and JP21K18660. This article is dedicated to Ahmed El Goresy who studied shocked meteorites and made excellent contributions to meteoritics and high-pressure mineral physics.

# References

Ahrens TJ (1987) Shock-wave techniques for geophysics and planetary physics, In Sammis CG, Henyey TL (Eds) Methods of Experimental Physics. Academic Press, New York, 185-235

Amsden A, Ruppel H, Hirt C(1980) Sale: a simplified ale computer program for fluid flows at all speeds. Tech. Rep. LA-8095 Report, Los Alamos National Laboratories

Bischoff A, Stöffler D (1992) Shock metamorphism as a fundamental process in the evolution of planetary bodies: Information from meteorites. Eur J Mineral 4:707-755

Buchwald VF (1975) Secondary structure of iron meteorite, Chapter 11, 125-136, in Handbook of iron meteorites, vol. 1, University of Hawaii.

Calister WD, Rethwisch DG (2000) Materials Science and Engineering: An Introduction. Eighth Edition, John Wiley & Sons, Inc.

Collins GS, Elbeshausen D, Davison TM, Wünnemann K, Ivanov BA, Melosh HJ (2016) iSALE-Dellen manual, Figshare, https://doi.org/10.6084/m9.figshare.3473690.v2

Cottrell AH, Aytekin VJ (1950) The flow of zinc under constant stress. Inst. Metals, 77, 389

Davidson AB (1940), The effect of annealing on Neumann bands. Electronic Theses and Dissertations. Paper 1705. https://doi.org/10.18297/etd/1705

Dunlop DJ, Özdemir Ö. (2007) Iron and Iron–Nickel, Treatise on Geophysics, in Treatise on Geophysics, 5, 277-336.

Farla RJM, Kokkonen H, Fitz Gerald JD, Barnhoorn A, Faul UH, Jackson I (2010) Dislocation recovery in fine-grained polycrystalline olivine. Phys Chem Minerals, DOI 10.1007/s00269-010-0410-3

Goldstein JI, Scott ERD, Chabot NL (2009) Iron meteorites: Crystallization, thermal history, parent bodies, and origin. Chem Erde-Geochem 69, 293–325.

Glover G and Stellars M, (1973) Recovery and recrystallization during high temperature deformation of α-Iron. Metallurgical Transactions, 4, 765-775.

Hilton CD, Bermingham KR, Walker RJ, Timothy J. McCoy TJ (2019) Genetics, crystallization sequence, and age of the South Byron Trio iron meteorites: New insights to carbonaceous chondrite (CC) type parent bodies. Geochim Cosmochim Acta 25, 217–228

Humphreys J, Rohrer GS, and Rollett A (2017), Recrystallization and Related Annealing Phenomena (The third edition). Elsevier, Netherlands, 691p. ISBN978-0-08-098235-9

Hwang H, Galtier E, Cynn H, Eom I, Chun SH, Bang Y, Hwang G, Choi J, Kim T, Kong M,

Kwon S, Kang K, Lee HJ, Park C, Lee JI, Lee Yongmoon, Yang W, Shim S-H, Vogt T, Kim Sangsoo, Park J, Kim Sunam, Nam D, Lee JH, Hyun H, Kim M, Koo T-Y, Kao C-C, Sekine T, Lee Yongjae (2020) Subnanosecond phase transition dynamics in laser-shocked iron. Sci. Adv.6, eaaz5132

Ivanov BA, Deniem D, Neukum G (1997) Implementation of dynamic strength models into 2-D hydrocodes: Applications for atmospheric breakup and impact cratering. Int. J. Impact Eng. 20, 411–430

Jain AV, Lipschutz ME (1968) Implications of shock effects in iron meteorites. Nature 220, 140-143.

Kruijer TS, Burkhardt C, Budde G, T. Kleine T (2017) Age of Jupiter inferred from the distinct genetics and formation times of meteorites, Proc Natl Acad Sci, 114 (2017), pp. 6712-6716

Li JCM (1962) Possibility of Subgrain Rotation during Recrystallization. J. Appl. Phys. 33, 2958‑2965

Marchi S, Durda DD, Polanskey CA, Asphaug E, Bottke WF, Elkins‑Tanton LT, et al. (2020) Hypervelocity impact experiments in iron‑nickel ingots and iron meteorites: Implications for the NASA Psyche mission. Journal of Geophysical Research: Planets, 125, e2019JE005927. https://doi.org/10.1029/2019JE005927

Michalak JT, Paxton HW (1961), Some Recovery Characteristics of Zone Melted Iron. Trans. AIME 221, 850–857

Miller GM, Stolper EM, Ahrens TJ (1991) The equation of state of molten komatiite 1. Shock wave compression to 36GPa. J Geophys Res 96 (B7):11831-11848

Monaghan, BJ and Quested, PN (2001) Thermal diffusivity of iron at high temperature in both the liquid and solid states. ISIJ international, 41, 1524-1528

Murr LE, Trillo EA, Bujanda AA, Martinez NE (2002a) Comparison of residual microstructures

associated with impact craters in fcc stainless steel and bcc iron target: the microtwin versus microband issue. Acta Mater 50: 121-131

Murr LE, Bujanda AA, Trillo EA, Martinez NE (2002b) Deformation twins associated with impact craters in polycrystalline iron target. J Mater Sci Lett 21:559-563

Prinz F, Argon AS, Moffatt WC (1982) Recovery of dislocation structures in plastically deformed copper and nickel single crystals, Acta Metall. 30, 821-830

Reisener RJ, Goldstein JI (2003) Ordinary Chondrite metallography: Part 1. Fe-Ni taenite cooking experiments. Meteoritics and Planetary Science 38, 1669-1678.

Rohde RW (1969) Dynamic yield behavior of shock-loaded iron from 76 to 573ºK. Acta Metallurgica, 17, 353-363

Scott ERD (2020) Iron Meteorites: Composition, Age, and Origin. Oxford Research Encyclopedias, Planetary Science. https://doi.org/10.1093/acrefore/9780190647926.013.206

Shinohara M (2002) Study of Development of a Compact Two- Stage Light Gas Gun and its Ignition System. Master of Science thesis, Institute of Fluid Science, Tohoku University.

Stöffler D, Keil K, Scott ERD (1991) Shock metamorphism of ordinary chondrites. Geochim Cosmochim Acta 55, 3845-3867.

Thompson SL (1990) ANEOS analytic equations of state for shock physics codes input manual. Sandia report, SAND89-2951·UC-404

Thompson SL, Lauson HS (1972) Improvements in the Chart D radiation-hydrodynamic CODE III: Revised analytic equations of state, *Rep. SC-RR-71 0714*, pp. 1– 119, Sandia Natl. Lab., Albuquerque, N. M.

Tomita M, Inaguma T, Sakamoto H, Ushioda K (2017) Recrystallization Behavior and Texture Evolution in Severely Coldrolled Fe-0.3mass%Si and Fe-0.3mass%Al Alloys. ISIJ

International 57: 921–928.

Uhlig HH (1955) Contribution of metallurgy to the origin of meteorites, Part II-The significance of Neumann bands in meteorites. Geochim Cormochim Acta 7: 34-42

Van Drunen G, Saimoto S (1971) Deformation and recovery of [001] oriented copper crystals. Acta Metall. 19, 213-221

Wasson JT (1967) The chemical classification of iron meteorites. I(Ge and Ga concentration in selected Fe meteorites used to determine quantization in terms of multiple parent body hypothesis and planetary fractionation processes). Geochim Cosmochim Acta, 31: 161-180

Wasson JT, Choi B-G, Jerde EA, Ulff-Møller F (1998) Chemical classification of iron meteorites: XII. New members of the magmatic groups. Geochim Cosmochim Acta, 62, 715-724

Watanabe T, Karashima S (1970) An analysis of high temperature creep in alpha iron based on the super jog mechanism. Transaction of the Japan Institute of metals 11: 159-165.

Wünnemann, K., Collins, G., Melosh, H., 2006. A strain-based porosity model for use in hydrocode simulations of impacts and implications for transient crater growth in porous targets. Icarus 180, 514‑527.

Yang J, Goldstein I, Scott ERD, Michael JR, Kotula PG, Pham T, McCoy TJ (2011) Thermal and impact histories of reheated group IVA, IVB, and ungrouped iron meteorites and their parent asteroids. Meteoritics & Planetary Science 46, 1227–1252. doi:10.1111/j.1945-5100.2011.01210.x

**Figure Legends**

Fig. 1. The two-stage light gas gun (installed at Institute of Fluid Science, Tohoku University)

used for impact experiments to simulate the formation of a Neumann band. A schematic image of the instrument components is also shown.

Fig. 2. A schematic image of the sample chamber with the heating and temperature measuring systems used for the two-stage light gas gun.

Fig. 3. The backscattered electron (BSE) image (a) and secondary electron image (SEI) (b) of the etched iron starting material.

Fig. 4. The images of craters formed by impacts at room temperature (Run 011), (a), 1100 K (Run 012), (b), and 670 K (Run 013) (c), and their polished cross sections.

Fig. 5. The optical and backscattered electron (BSE) images of the texture of the polished section adjacent to the crater surface within 3-mm depths of the sample impacted at room temperature (Run 011). Its high magnification secondary electron images (SEI) are also shown.

Fig. 6. The optical and backscattered electron (BSE) images from the top to the bottom of the target sample of the room temperature impact (Run 011) and its magnified secondary electron images (SEI). The twins are observed in the images.

Fig. 7. The optical and backscattered electron (BSE) images of the polish section and the magnified secondary electron images (SEI) of the different positions of the sample recovered from the impact at 1100 K (Run 012).

Fig. 8. The optical and backscattered electron (BSE) images of the sample impacted at 670 K (Run 013) and the secondary electron images (SEI) magnified locally.

Fig. 9. The BSE image from the top to the bottom of the recovered sample annealed with the induction furnace at 1070 K for 10 min, and its locally magnified SEI images (Run 011H).

Fig. 10. The results of numerical simulations on distributions of the maximum temperature (left) and pressure (right) in the iron targets during the impacts at 300 K (a), 1100 K (b), and 670

K (c) corresponding to the experimental conditions of Run 011, 012, and 013. The color contour on the left side indicates the peak temperature, Kelvin, and that on the right side is the peak pressure in GPa.

Fig. 11. The results of the iSALE-2D simulation on distributions of the peak pressure (GPa) and temperature (K) along the center axis of the target. The distribution of peak pressure (a) and temperature (b) in the room temperature impact (corresponding to Run 011), the peak pressure (e) and temperature (d) in the impact at 670 K (corresponding to Run 013), and the peak pressure (e) and temperature (f) in the 1100 K impact (corresponding to Run 012). The distance from the crater is shown as the depth in mm.

Fig. 12. (a) The twin density from the top to the bottom of the samples recovered from the impacts at room temperature (Solid squares, Run 011) and 670 K (Open circles, Run 013). The density after annealing at 1070 K (Open squares, Run011H) for 10 min is also shown. The twin distribution is expressed as a linear relation, N (cm$^{-2}$) = 4.14 × 10$^5$ – 2.34 × 10$^4$ × D (mm) (R = 0.908) for the room temperature impact, and N (cm$^{-2}$) = 4.45 × 10$^5$ – 2.05 × 10$^4$ × D (mm) (R = 0.936) for the impact at 670 K, where N is the density of twins in a unit area of 1 cm$^2$, and D is the depth of the sample from the crater surface in mm. (b) The relation between the twin density and pressure (GPa) that was deduced by the iSALE-2D simulation. The relation can be expressed by N (cm$^{-2}$) = –6.30 × 10$^4$ + 4.11 × 10$^4$ × P (GPa) (R = 0.905). All data for room temperature (solid squares) and 670 K (open circles) are included for the fitting except a data point (*) at 13 GPa in which the twin density close to the crater may be reduced by increasing pressure and temperature.

Fig. 13. The effect of time (a) and temperature (b) for annealing of twins assuming n = 2 and 3 for decreasing rate expressed by an empirical n-th order recrystallization kinetics expression, d$\rho$/dt = –k$\rho^n$, where $k$ is the rate constant for recrystallization (see the text for more details). (a) Decrease in twin density by annealing during time intervals from one minute to one hour at 1070 K and various depths of the sample for n = 2 (blue curves) and for n = 3 (red dotted curves). The experimental data for Run 011 and Run 011H (10-min annealing at 1070 K) are shown as black open and red open circles respectively this figure. (b) The change in twin density by annealing at various temperatures and durations from one minute to one hour for n = 2 (blue curves) and n = 3 (red curves). The initial twin density before annealing was taken as $3.5 \times 10^5$ cm$^{-2}$, and the activation energy $E_a$ of 250 kJ/mol was adopted for the recrystallization process of twins.

Supplementary information

Fig. S1. The piston made of high-density polyethylene (HDPE) used for the first stage of compression. A piston before the impact experiment (a) and that after the experiment (b).

Fig. S2. The projectile used for the impact experiments.

Fig. S3. The target sample (iron rod) in a molybdenum capsule.

Table 1. Conditions for the impact and annealing experiments.

| | Second diaphragm | Piston | Projectile | | | Target | | |
|---|---|---|---|---|---|---|---|---|
| | | | Material | weight (g) | Velocity (km/s) | Material | Size (D×H), mm | Temperature, K |
| Run 011 | Mylar, 0.5mm | HDPE | PC +iron | 2.46 | 1.53 | Iron | 38×20 | 300 |
| Run 012 | Mylar, 0.5mm | HDPE | PC +iron | 2.43 | 1.51 | Iron | 38×30 | 1100 |
| Run 013 | Mylar, 0.5mm | HDPE | PC +iron | 2.43 | 1.47 | Iron | 38×30 | 670 |
| Run 011H* | - | - | - | - | - | Iron | 18×2×1** | 1070 |

HDPE, High Density Polyethylene: PC, Polycarbonate, *, annealing experiment, **Height ×Width ×Thickness

Table 2. Number Density (cm$^{-2}$) of Neumann bands with depths for the collision at room temperature (Run 011), 670 K (Run 013), and after annealing at 1070 K for 10 minutes (Run 011H). *Pressure was estimated by impact simulation (see text).

| Depths (mm) | Number in 400μm×300μm | Mean grain size (μm) | Std dev. (μm) | Number density in one grain (cm$^{-2}$) | Pressure (GPa)* |
|---|---|---|---|---|---|
| Run 011 Collision at 300 K | | | | | |
| 2.9 | 361 | 134 | 26 | $3.0(1.2)\times10^5$ | 12.7 |
| 6 | 384 | 125 | 30 | $3.2(1.5)\times10^5$ | 9.26 |
| 7.7 | 269 | 102 | 24 | $2.2(1.1)\times10^5$ | 7.65 |
| 9 | 314 | 129 | 33 | $2.6(1.3)\times10^5$ | 6.71 |
| 11.6 | 131 | 115 | 38 | $1.1(0.7)\times10^5$ | 5.41 |
| 13.8 | 87 | 113 | 37 | $7.3(4.7)\times10^4$ | 4.07 |
| Run 013. Collision at 670 K | | | | | |
| 4.9 | 421 | 154 | 25 | $3.5(1.1)\times10^5$ | 10.2 |
| 6.4 | 297 | 157 | 34 | $2.5(1.1)\times10^5$ | 8.8 |
| 9.3 | 382 | 206 | 25 | $3.2(0.8)\times10^5$ | 7 |
| 15.4 | 184 | 200 | 21 | $1.5(0.3)\times10^5$ | 3.89 |
| 20 | 11 | 170 | 42 | $9.2(4.5)\times10^3$ | 2.45 |
| Run 011H. Annealing at 1070 K for 10 minutes | | | | | |
| 2.7 | 40 | 68 | 15 | $3.3(1.5)\times10^4$ | - |
| 7 | 65 | 128 | 37 | $5.4(3.1)\times10^4$ | - |
| 9.2 | 55 | 137 | 35 | $4.6(2.3)\times10^4$ | - |
| 11.8 | 53 | 143 | 50 | $4.4(3.1)\times10^4$ | - |
| 16.4 | 45 | 126 | 40 | $3.8(2.4)\times10^4$ | - |

Table 3. Models and parameters used for simulation with the iSALE 2D code.

| Parameters used for simulation | ARMCO iron (Johnson and Cook, 1985) |
|---|:---:|
| Analytical EOS (ANEOS) | Thompson (1990) |
| Melting temperature | 1811 |
| Specific heat (J/kgK) | 452 |
| Strength model | Johnson-Cook (JNCK) model |
| JNCK parameters | |
| A(MPa) | 175 |
| B(MPa) | 380 |
| N | 0.32 |
| C | 0.06 |
| M | 0.55 |
| $T_{ref}(K)$ | 298 |

Table 4. Calculation settings used in the numerical simulations with the iSALE*.

| Computational geometry | Cylindrical coordinate |
|---|---|
| Number of computational cells | |
|    R direction | 330 |
|    Z direction | 600 |
| Number of cells per projectile radius | |
|    Radial direction | 57 |
|    Vertical direction | 25 |
| Number of cells per target radius | |
|    Radial direction | 271 |
|    Vertical direction | 214 (670 K, 1100 K) <br> 143 (300 K) |
| Artificial viscosity $a_1$ | 0.24 |
| Artificial viscosity $a_2$ | 1.2 |
| Impact velocity (km s$^{-1}$) | The same archived in the experiments |
| High-speed cutoff | 2-fold impact velocity |
| Low-density cutoff (kg m$^{-3}$) | 100 |

*Detailed descriptions of the parameters can be found in the iSALE manual (Collins et al., 2016).

Figure 1 (Ohtani)

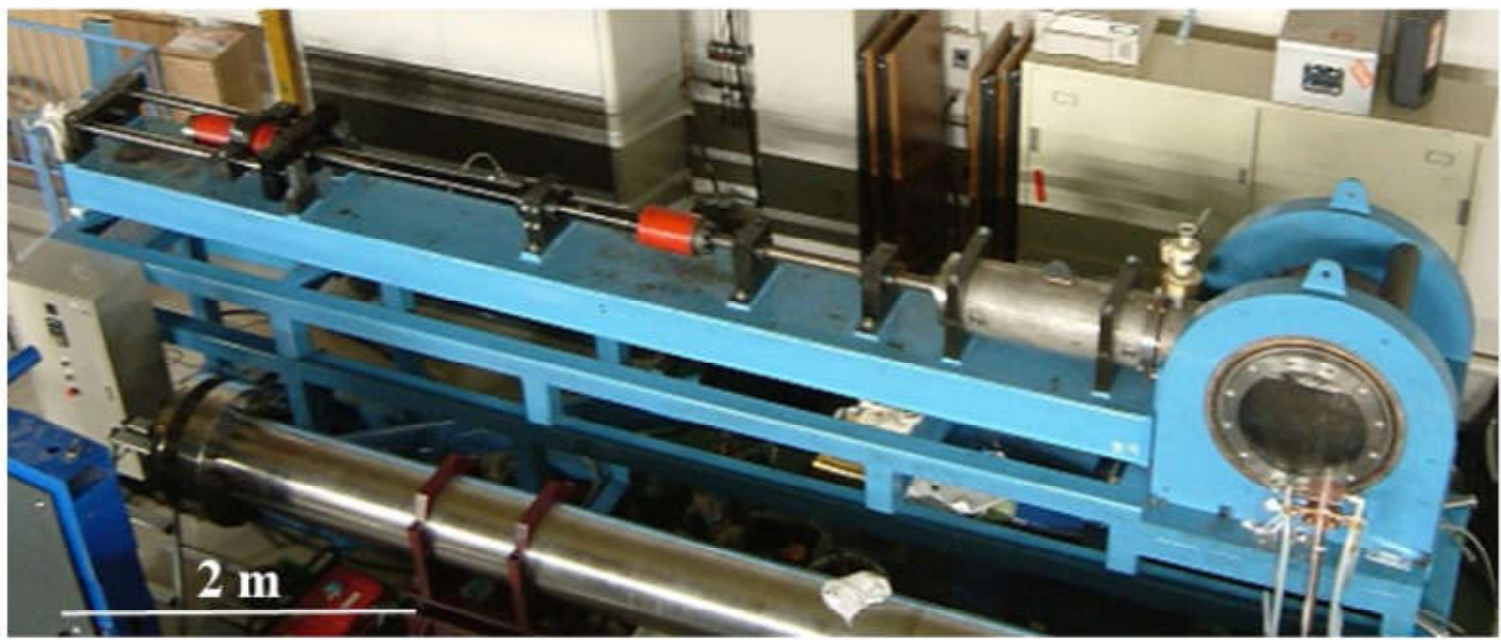

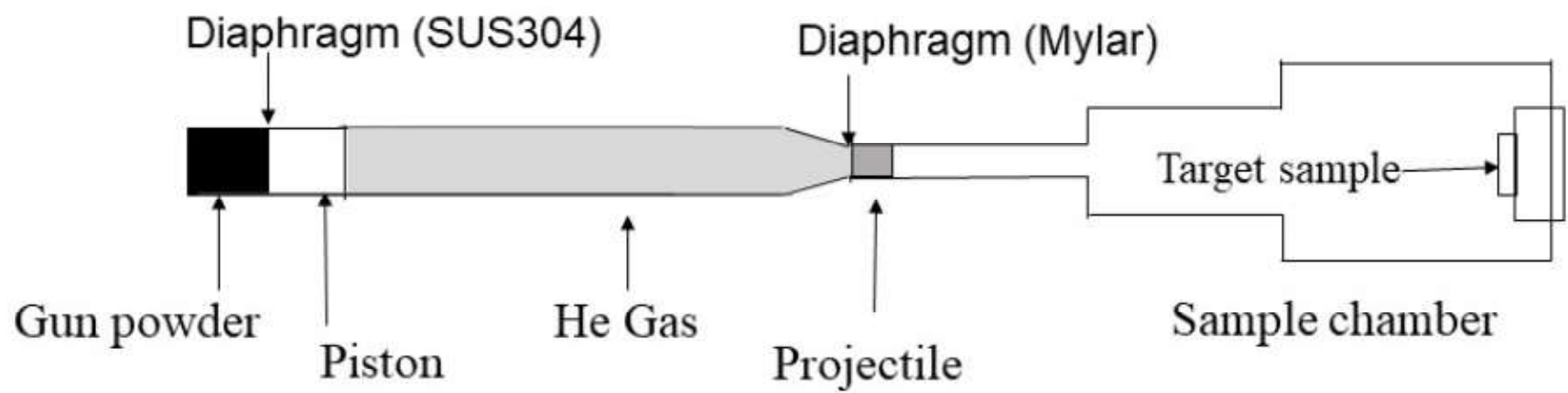

Figure 2 (Ohtani)

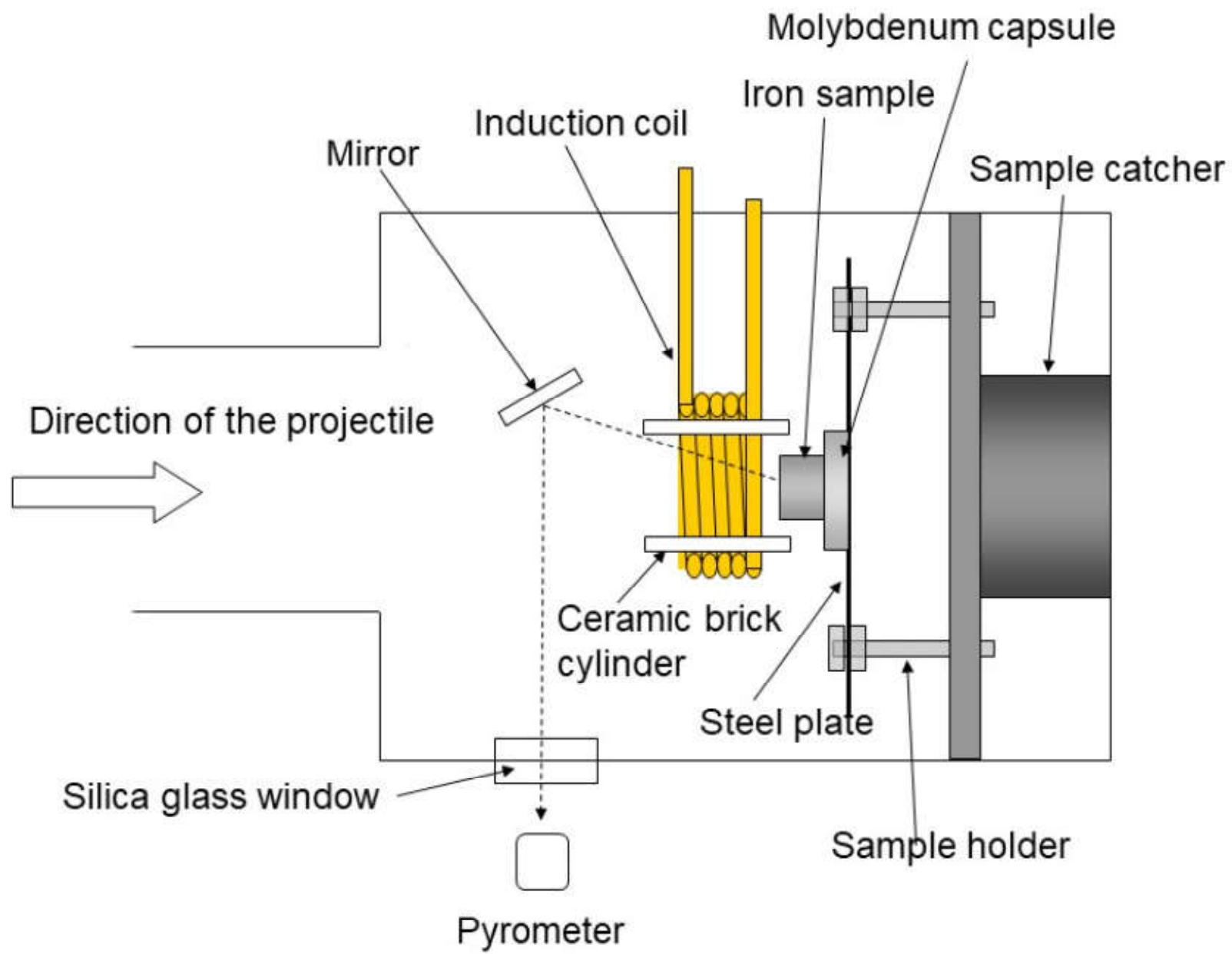

Figure 3 (Ohtani)

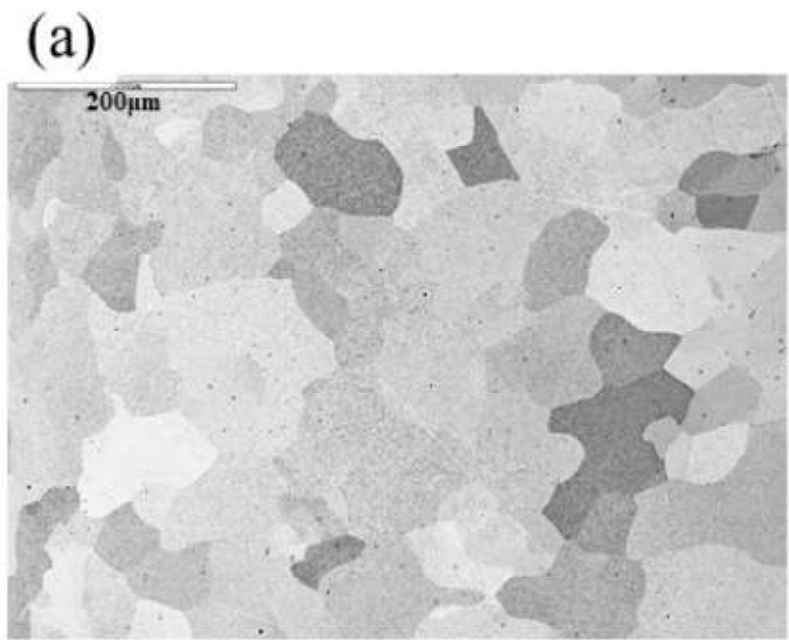 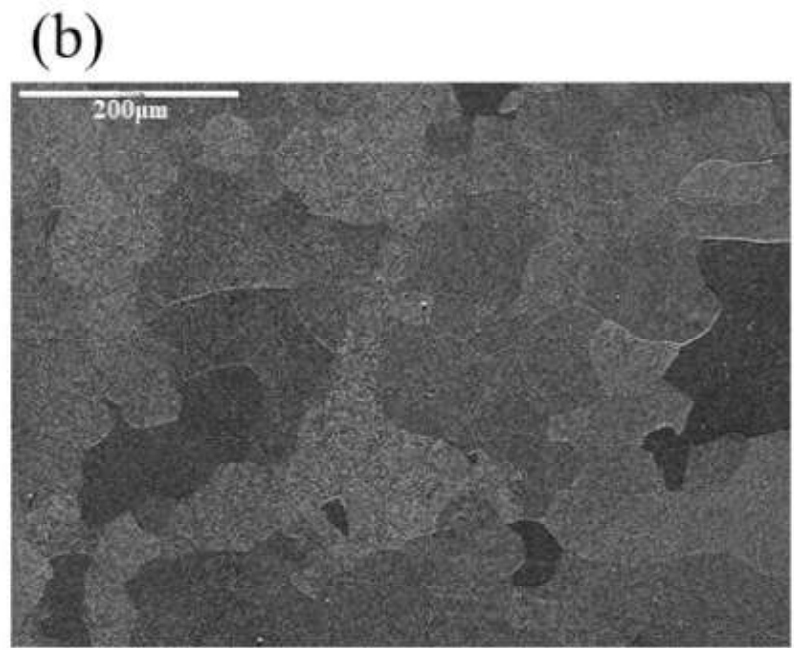

Figure 4 (Ohtani)

(a) 300 K (Run 011)  (b) 1100 K (Run 012)  (c) 670 K (Run 013)

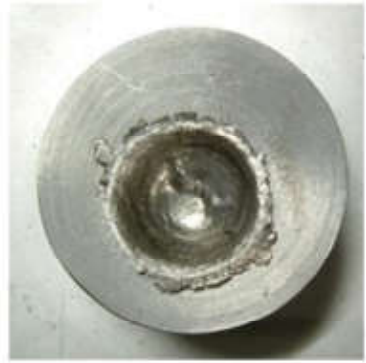
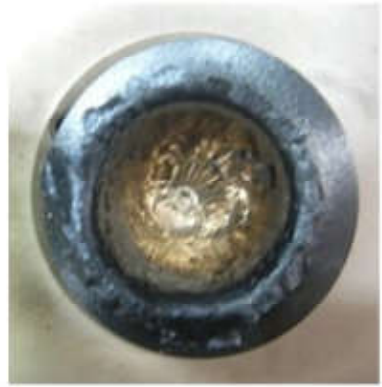
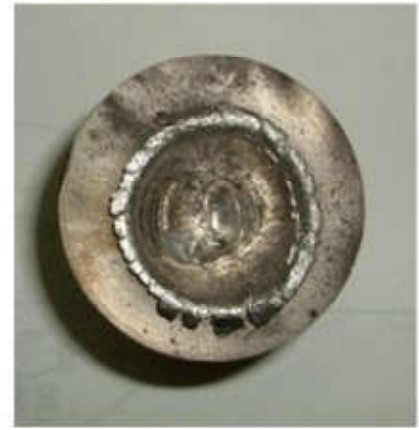
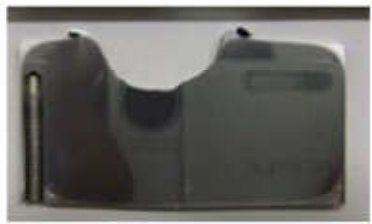
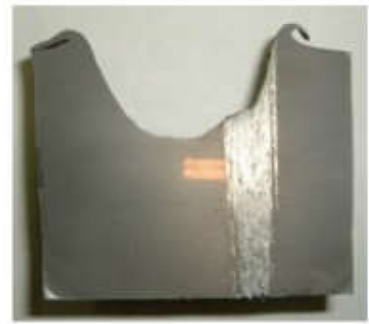
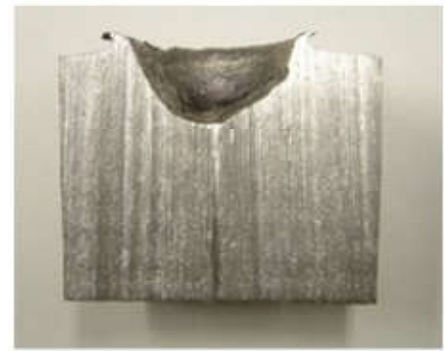

38mm    38mm    38mm

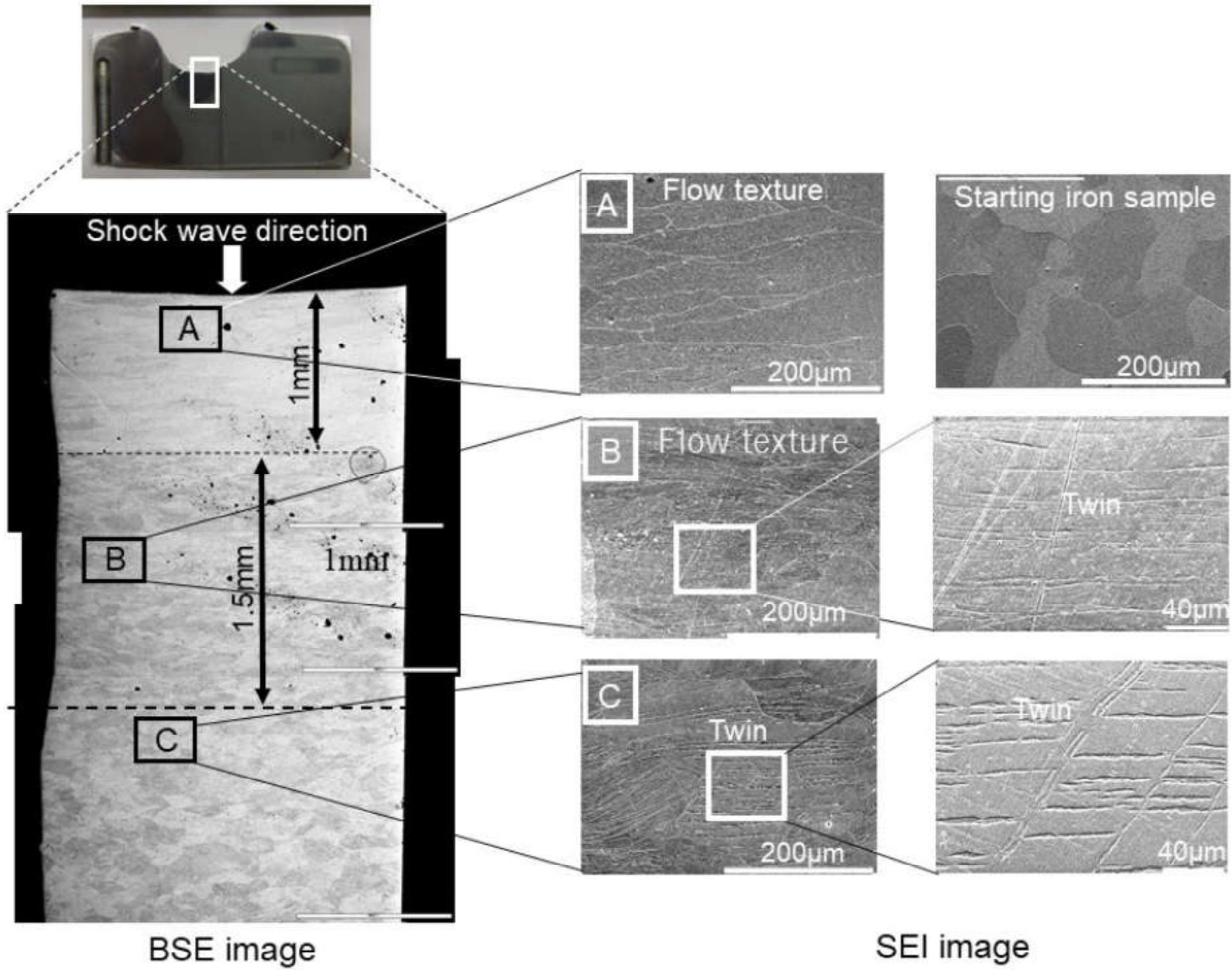

BSE image

SEI image

Figure 6 (Ohtani)
Run 011

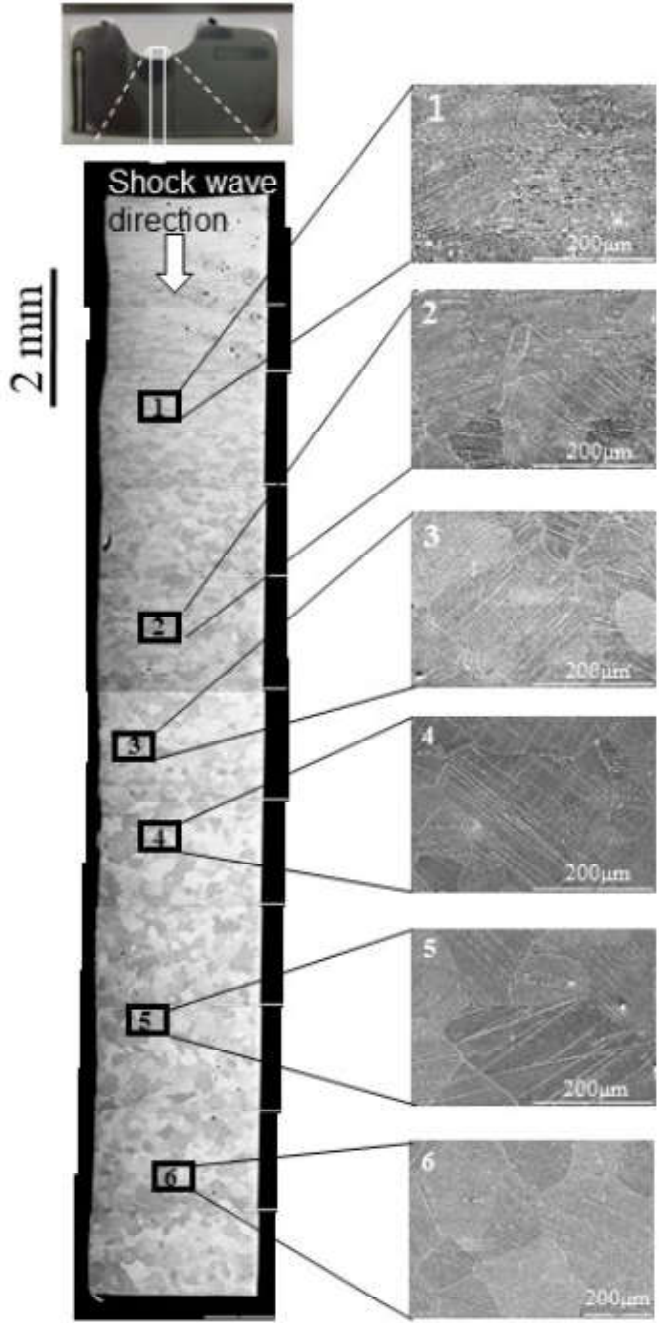

Room temperature impact

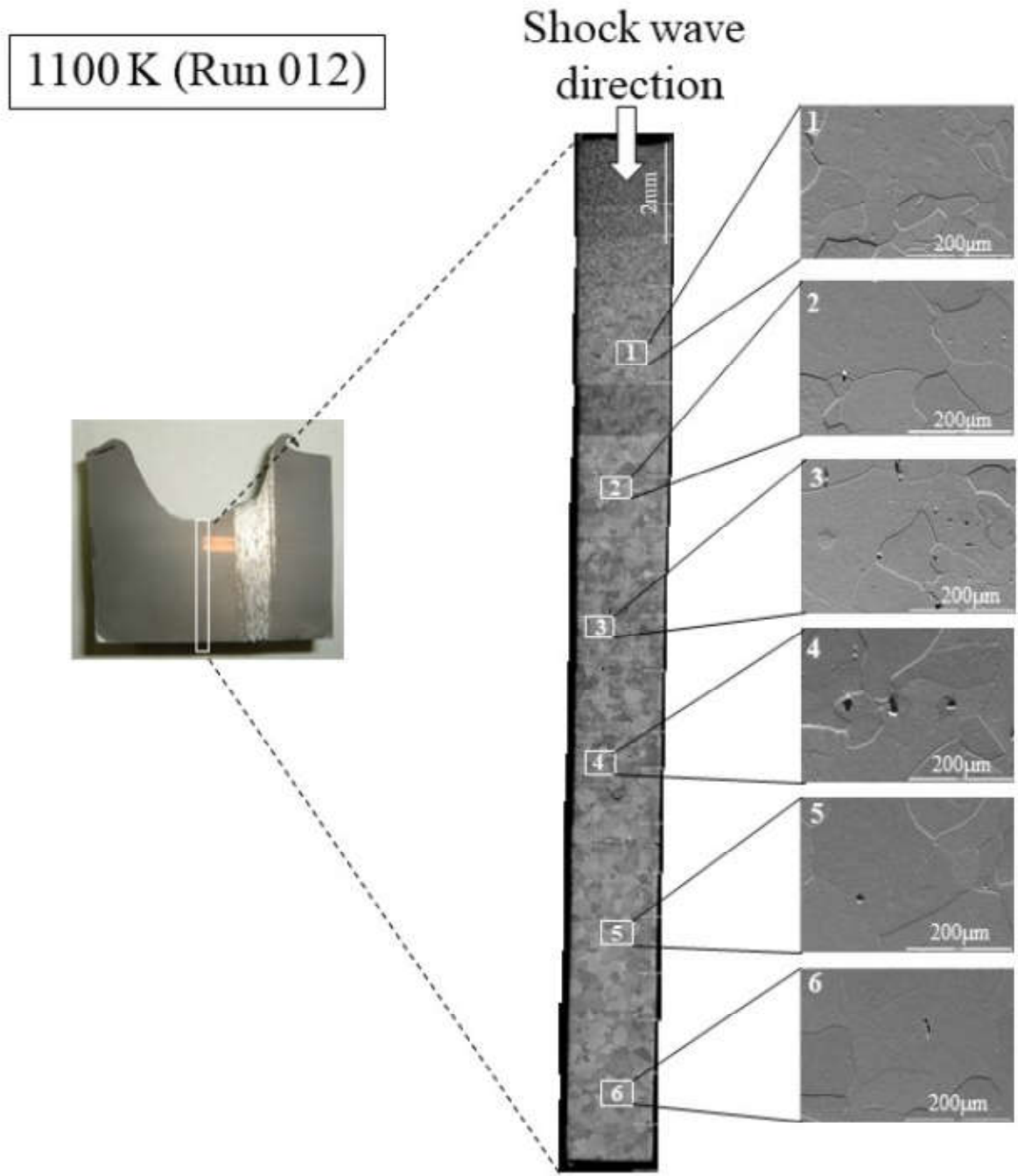

Figure 7 (Ohtani)

670 K (Run 013)

Figure 8 (Ohtani)

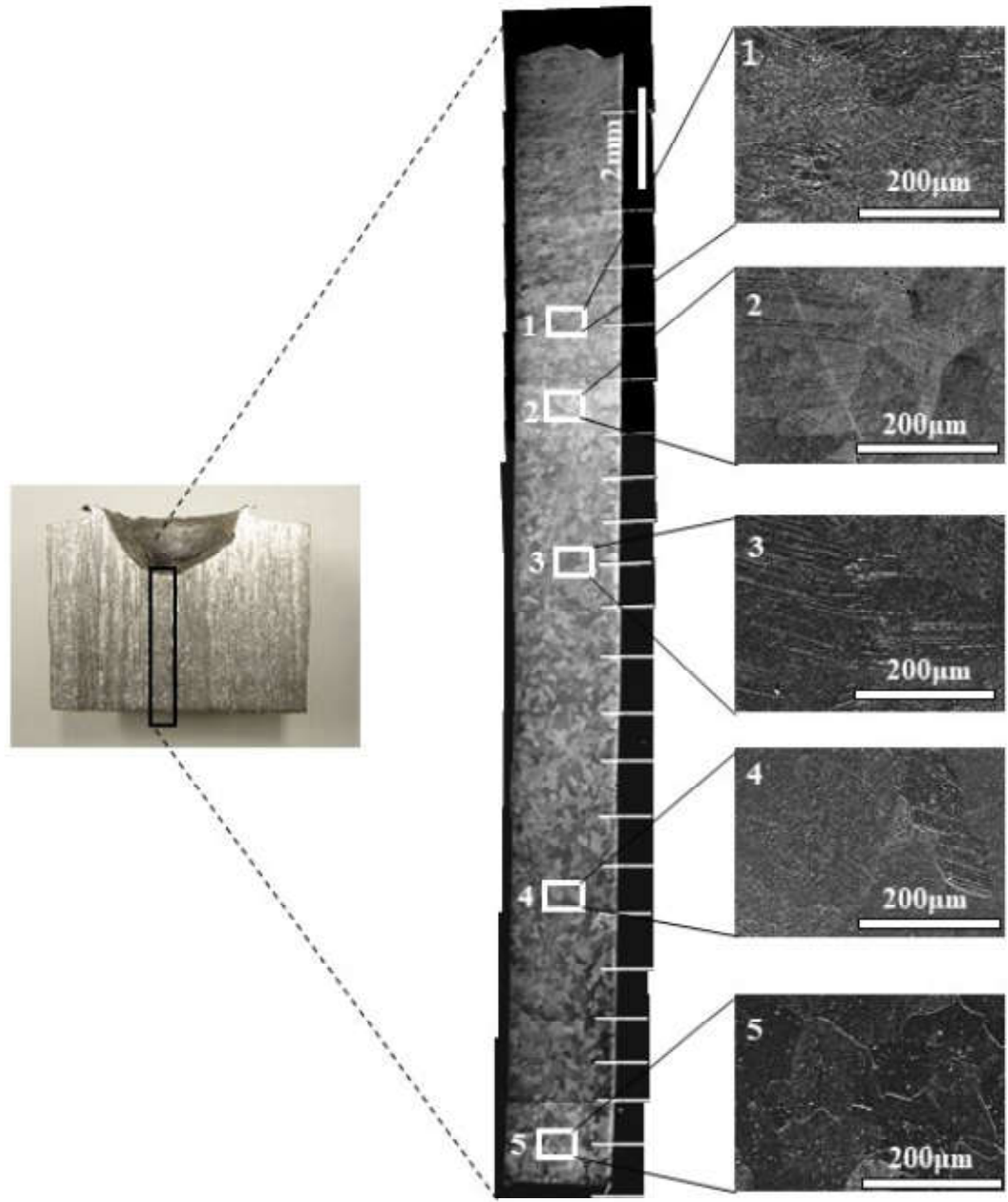

Figure 9 (Ohtani)

1070 K annealing for 10 minutes (Run 011H)

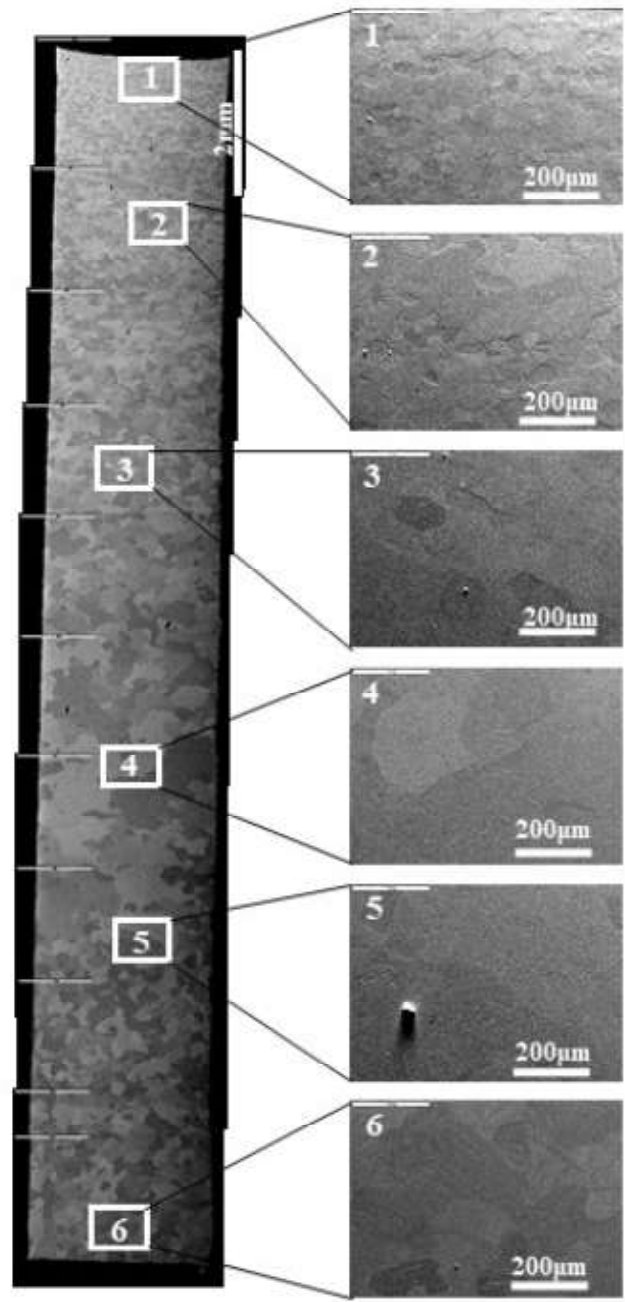

Figure 10(a) Run 011 Collision at 300 K (ARMCO iron)

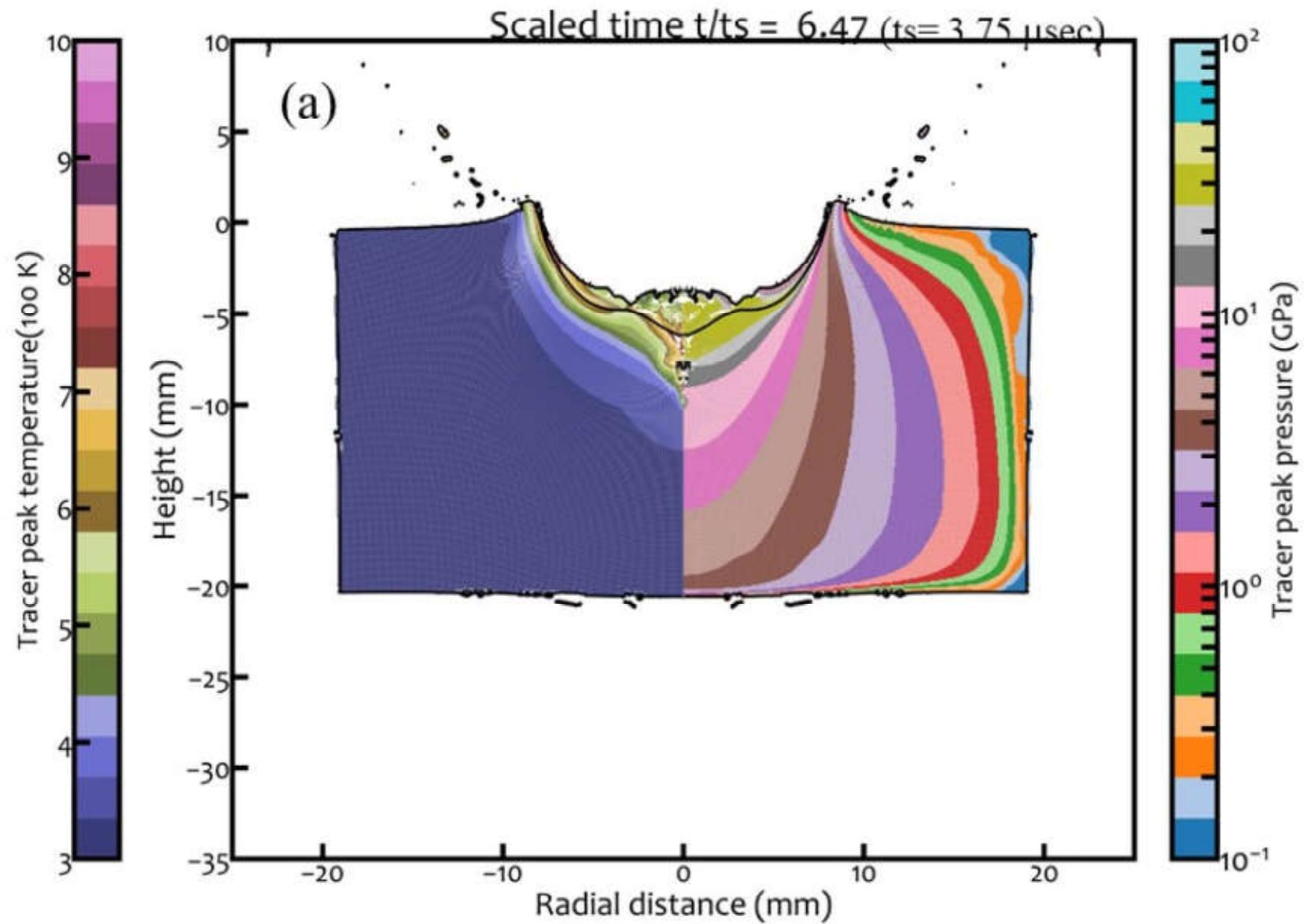

Figure 10(b)    Run 012  Collision at 1100 K (ARMCO iron)

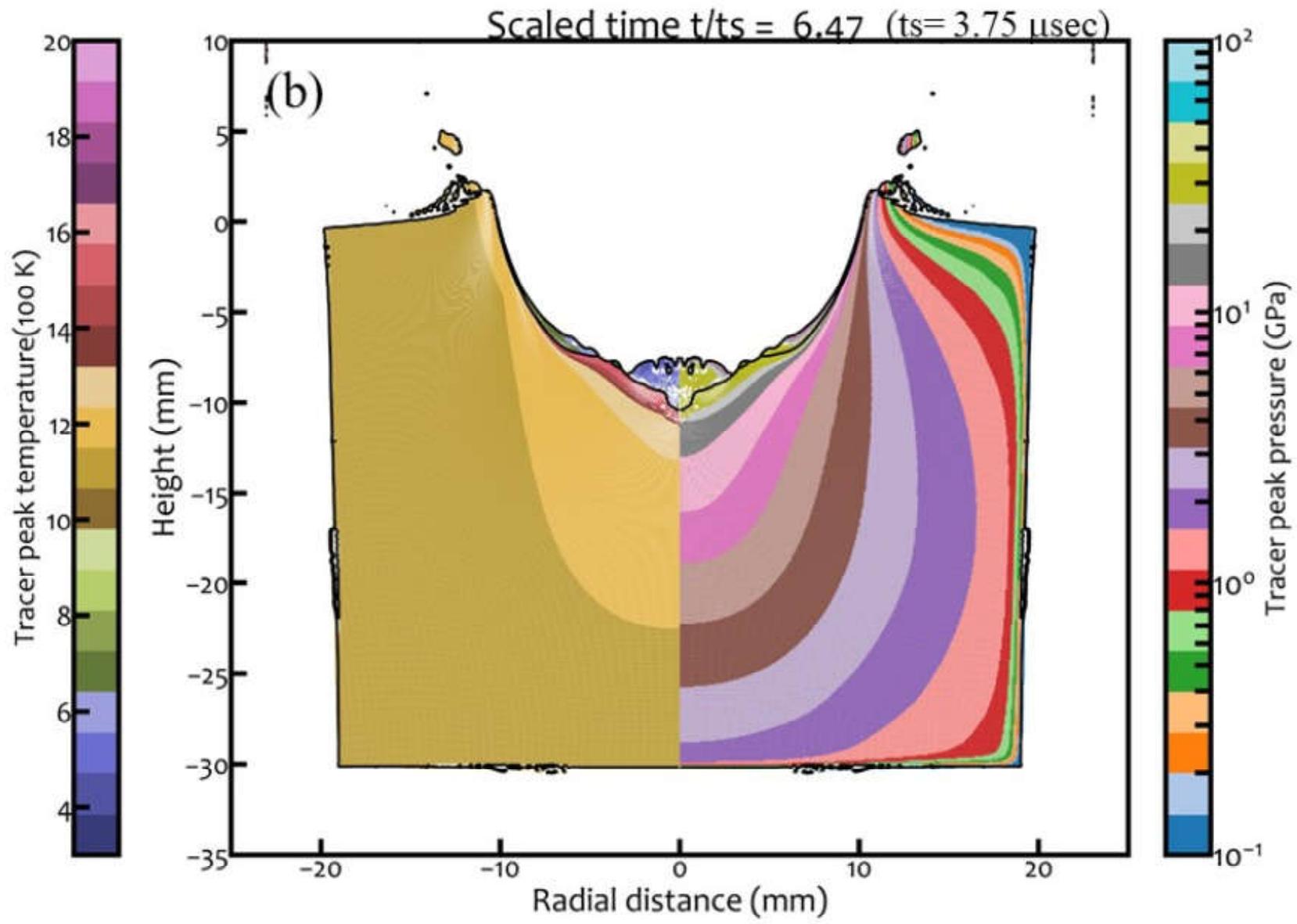

Figure 10(c)  Run 013  Collision at 670 K (ARMCO iron)

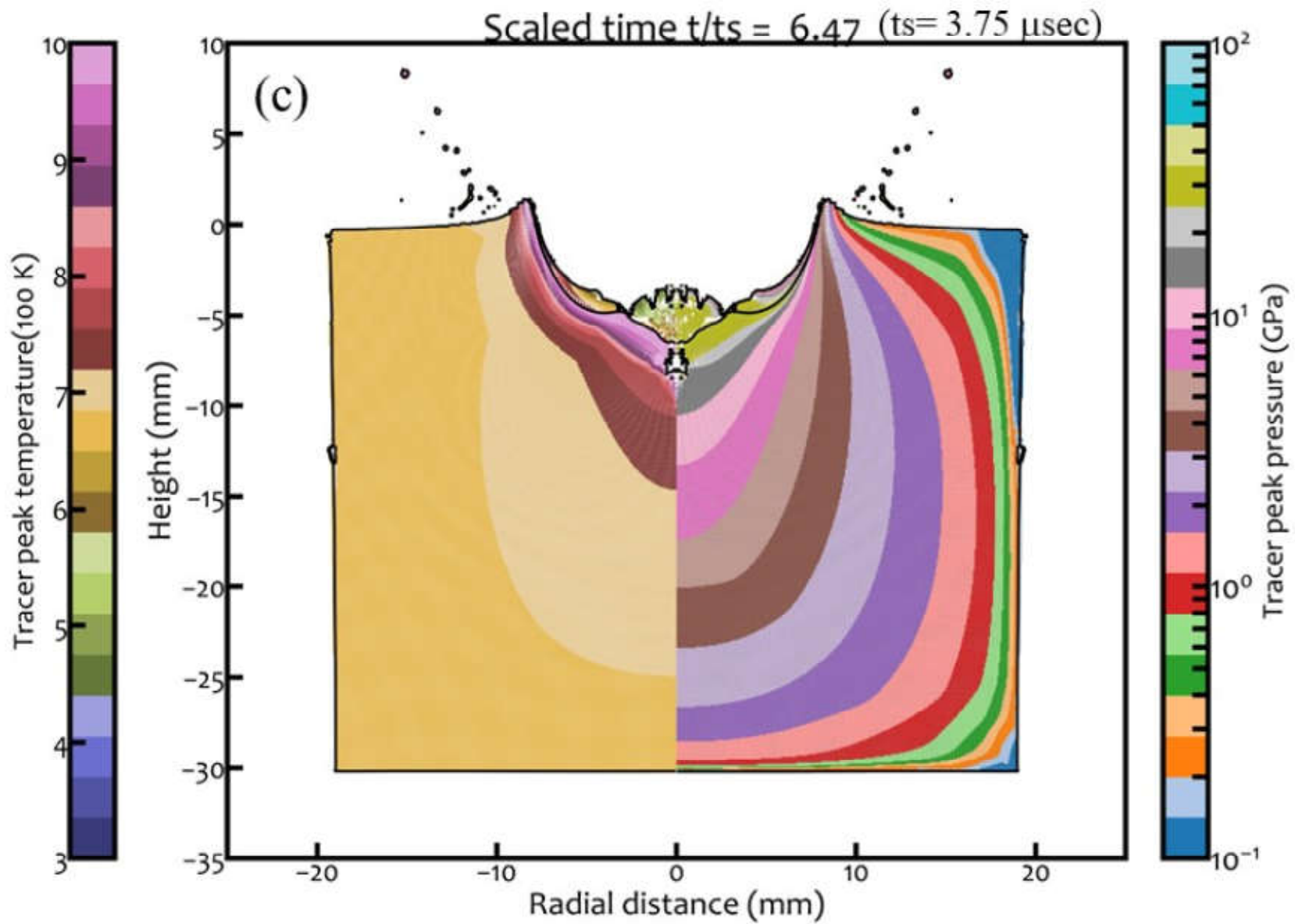

Figure 11

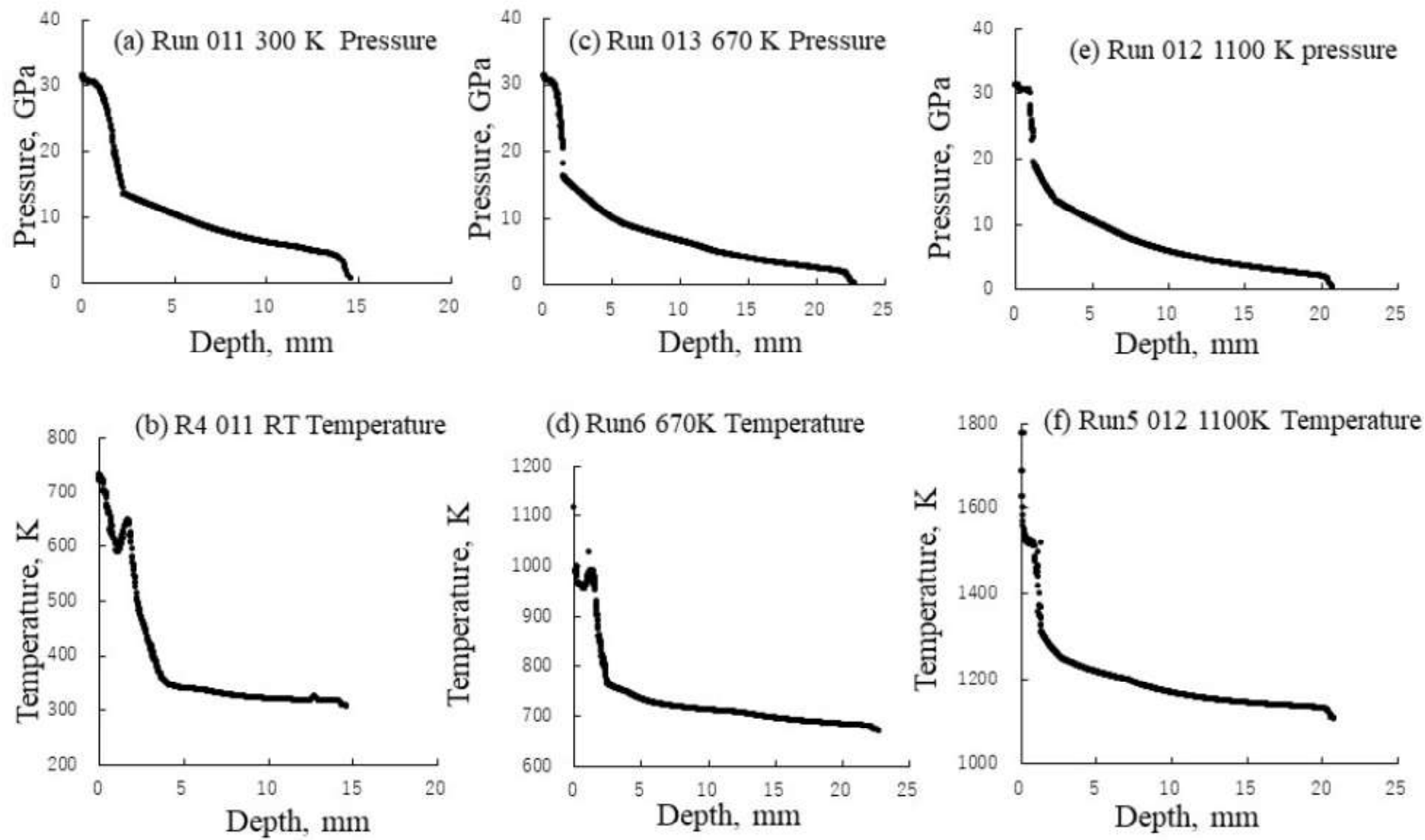

Figure 12 (Ohtani)

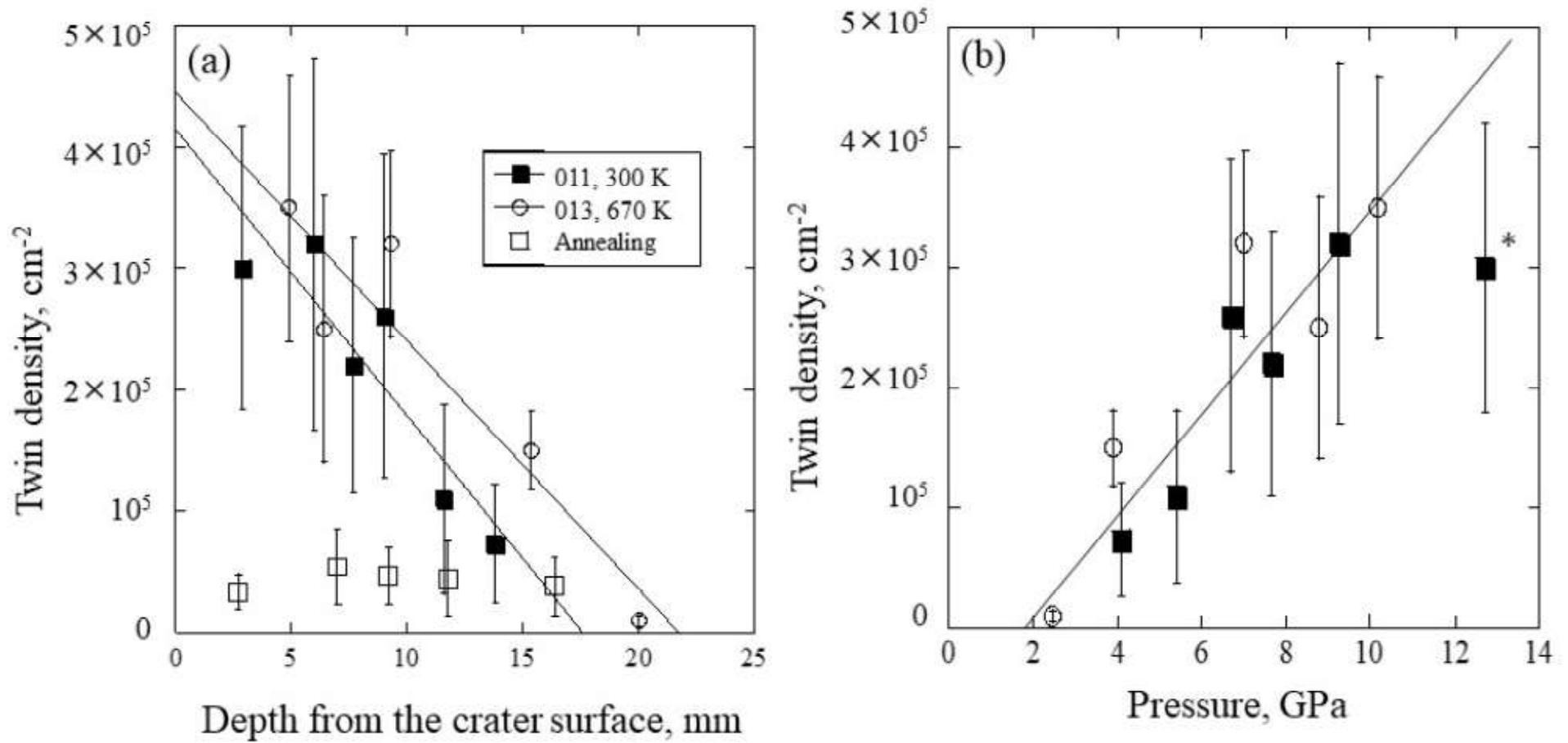

Figure 13

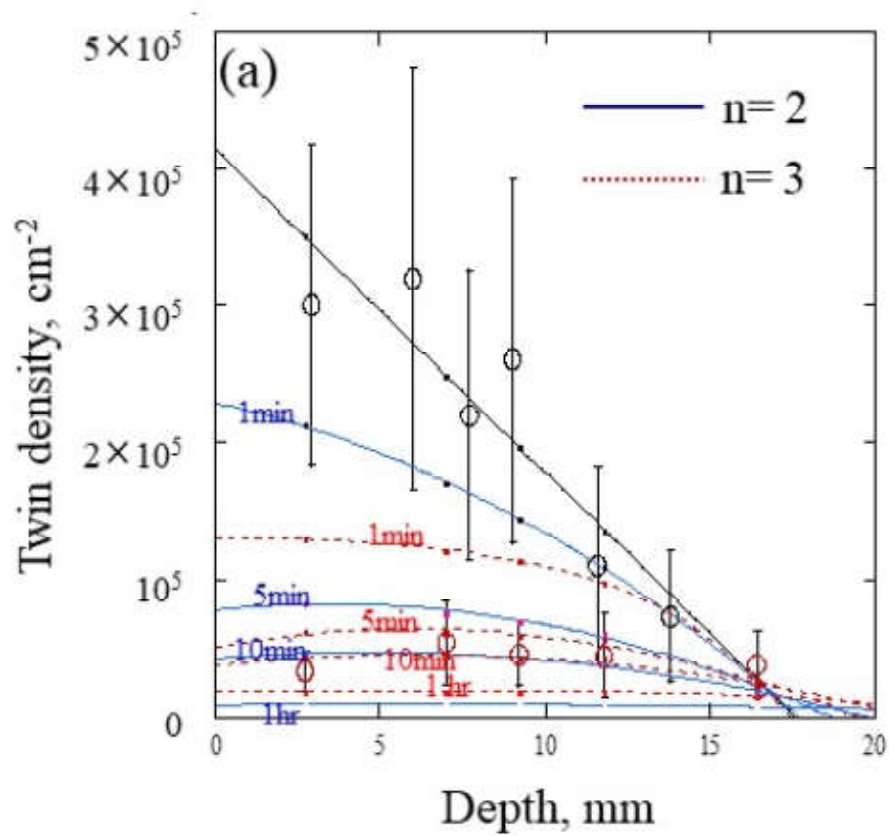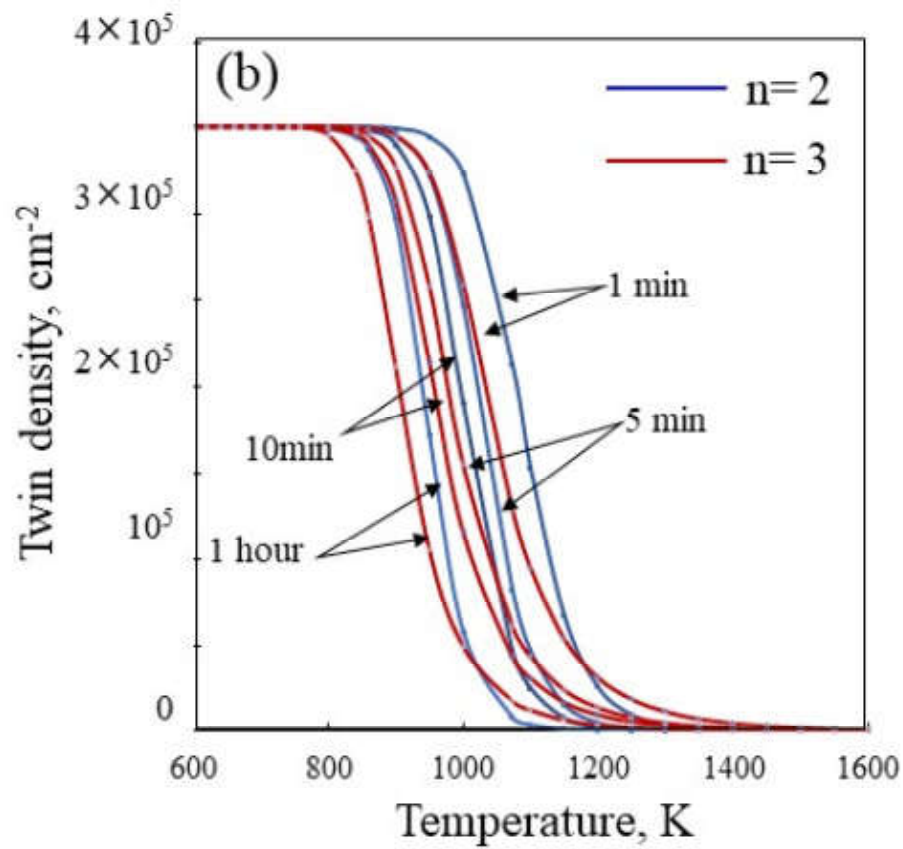

Figure S1 (Ohtani)

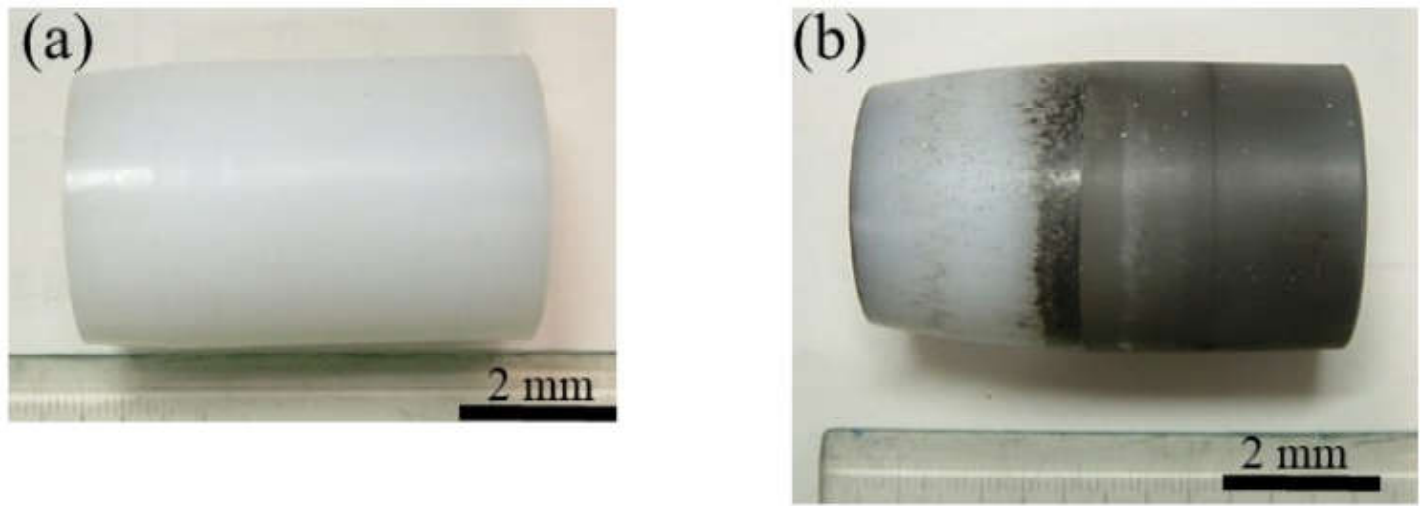

Figure S1. The piston made of high density polyethylene (HDPE) was used for the first stage compression. Pistons before (a) and after (b) the collision experiment.

Figure S2 (Ohtani)

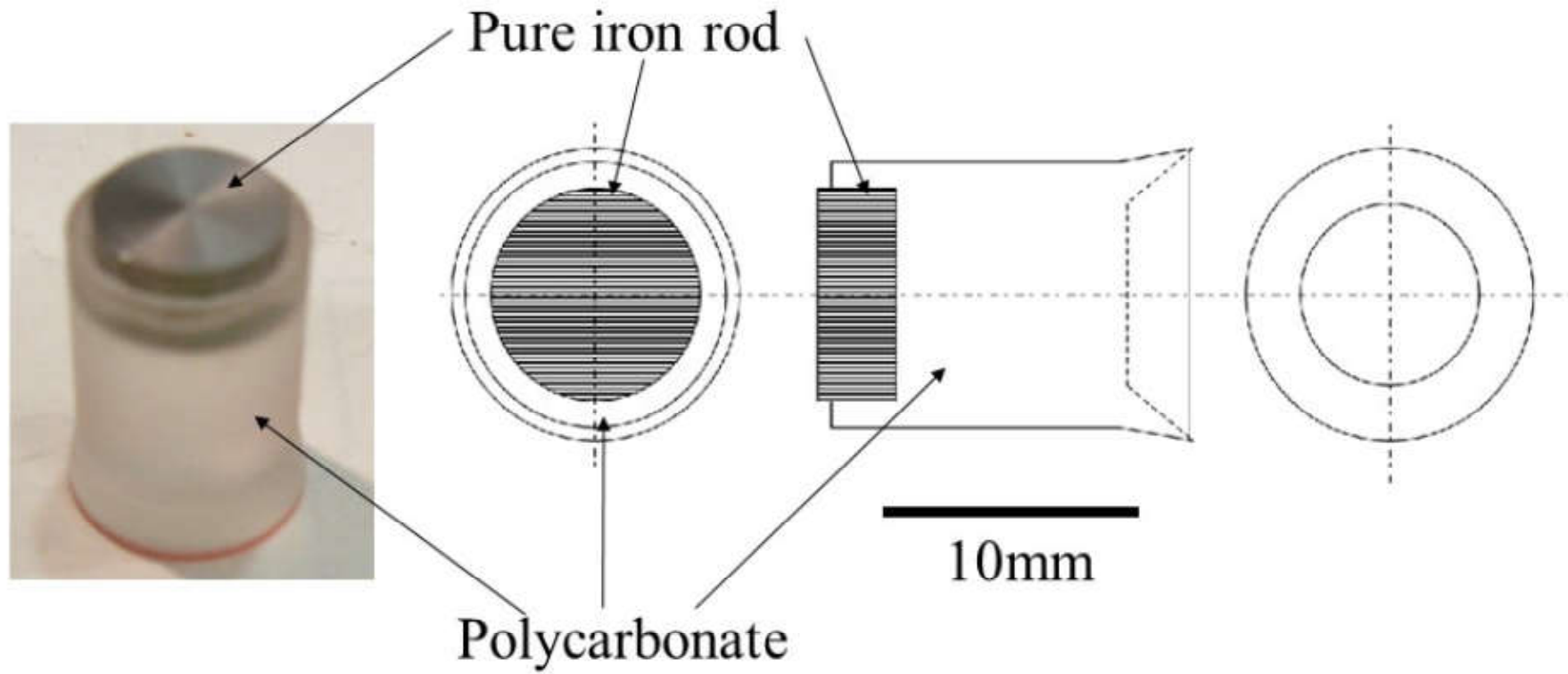

Figure S2. Projectile used for the collision experiments.

Figure S3 (Ohtani)

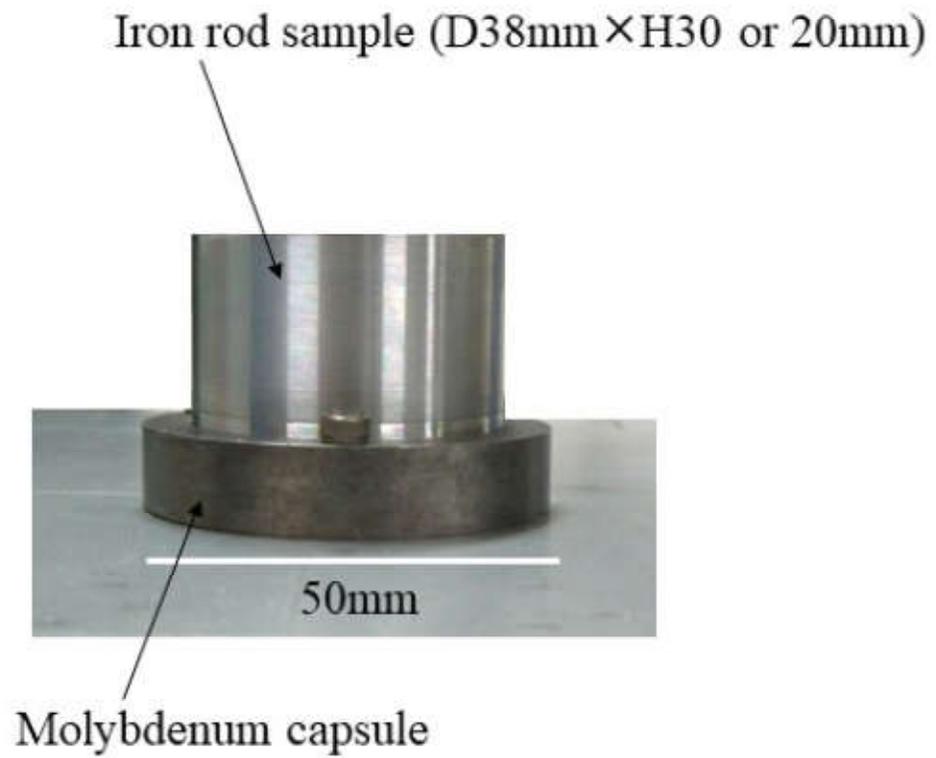

Iron rod sample (D38mm×H30 or 20mm)

50mm

Molybdenum capsule